The effects of non-stoichiometric lattice structure and secondary phase α-$Fe_2O_3$ on modifying the magnetic, electrical, and magneto-electric properties of Eu doped $BaFe_{12}O_{19}$ hexaferrite

Gara Kishor[a], R. N. Bhowmik[a]*, P. D. Babu[b]

[a] Department of Physics, School of Physical, Chemical and Applied Sciences, Pondicherry University, R. V Nagar, Kalapet-605014, Puducherry, India.

[b] UGC – DAE Consortium for Scientific Research Mumbai Centre, Bhabha Atomic Research Centre, Mumbai-400085, India.

*Corresponding author: rnbhowmik.phy@pondiuni.edu.in

Abstract

The Ba-atoms in $BaFe_{12}O_{19}$ were substituted by Eu-atoms to prepare the material of composition $Ba_{1-x}Eu_yFe_{12}O_{19}$. The atomic ratio of Eu/Ba were varied from 0 to 0.33 to control the secondary phase α-$Fe_2O_3$ and properties of M-type hexaferrite (magneto-plumbite) structure. Rietveld refinement of the X-ray diffraction patterns suggested the occupancy of the Eu atoms at the sites of Fe atoms instead of the Ba atoms. This promoted the formation of the secondary phase α-$Fe_2O_3$ (hematite). Raman spectra provided a complimentary support for a systematic increase of the hematite phase on increasing the Eu/Ba ratio. X-ray photoelectron spectroscopy evidenced the presence of oxygen deficiency and surface defects. The charge state of Ba and Fe/Eu ions are maintained close to +2 and +3, respectively. Saturation magnetization of the system decreased from 84.44 emu $g^{-1}$ to 57.68 emu $g^{-1}$ at 3 K and from 67.09 emu $g^{-1}$ to 43.42 emu $g^{-1}$ at 300 K with the increase of AFM hematite phase fraction, Unlike remarkable enhancement of the coercivity increases 914 Oe to 3656 Oe at 3 K and from 1996 to 4862 Oe at 300 K. Electrical conductivity of the system increased from $10^{-11}$-$10^{-10}$ $Sm^{-1}$ at 30 K to $10^{-5}$ -$10^{-4}$ $Sm^{-1}$ at 300 K and confirmed semiconductor behavior. Temperature-dependent electrical conductivity analysis reveals the Efros-Shklovsky-type variable-range hopping conduction mechanism at the low-temperature region, whereas the Mott variable range hopping conduction mechanism is applicable near to room temperature. Electric field dependent charge transport properties obeyed the Schottky emission mechanism, yielding electrode and sample interface barrier height of 0.27 – 0.74 eV, and the Poole–Frenkel emission mechanism, providing trap barrier height of 0.18 - 0.40 eV. The results demonstrate that non-stoichiometric lattice structure can be an effective strategy to suitably modify the magnetic and electrical properties in M-type Ba hexaferrite material.

## 1.Introduction

The M-type barium hexaferrites ($BaFe_{12}O_{19}$) highly promising for applications in permanent magnets, microwave absorbers, and data storage devices [1]. This is due to the fact that $BaFe_{12}O_{19}$ (BM) showed high uniaxial magneto-crystalline anisotropy (~ $10^5$ J $m^{-3}$), high magnetization (60–80 emu $g^{-1}$), moderate coercivity (3–7 kOe), high curie temperature (720–730 K), electrical conductivity (~ $11.7 \times 10^{-9}$ S $cm^{-1}$) and stable chemical structure [2]. The unit cell in the magneto-plumbite structure consists of two formula units of $BaFe_{12}O_{19}$, which are stacked in RSR*S* (R - rhombohedral unit and S-spinel unit) sequence along the c-axis. The * represents 180º rotation of the respective blocks along c-axis. The high spin $Fe^{3+}$ (5 $\mu_B$) ions are distributed over five crystallographically distinct coordination sites, viz., three octahedral (1a, $4f_2$, 12k) sites, one tetrahedral ($4f_1$) site, and one trigonal bipyramidal (2b) site [1, 3]. The substitution of transition metal, rare earth metal and other elements in the BaFe12O19 structure, either at the Fe-sites or Ba-sites, showed a great impact on modifying the structural, optical, electrical, magnetic and magneto-electric properties [4, 5]. Generally, rare earth (RE) ions exhibit a strong spin–orbit coupling, large orbital moments and localized 4f magnetic moments [6 - 9]. Among the RE elements, Europium (Eu) ion can exhibit in multi valence charge states ($Eu^{2+}$ and $Eu^{3+}$) [10]. The half-filled $4f^7$ electronic configuration of $Eu^{2+}$ ion can contribute a large spin magnetic moment (theoretical value ~ 7.94 $\mu_B$ and experimental value ~ 6.2 - 6.6 $\mu_B$) [11] and zero orbital angular moment contribution (J = S = 7/2, L = 0). On the other hand, theoretical prediction of non-magnetic ground state ($^7F_0$, J = 0, L = S = 3) in partially filled $Eu^{3+}$ ($4f^6$) ion differs from the experimental magnetic moment typically 3.4 – 3.6 $\mu_B$ at room temperature. This is due to the fact that thermal energy is sufficient to overcome a small energy gap between the ground state ($^4F_0$) and the low-lying excited states ($^7F_1$) and increases the spin population at the $^7F_1$ states to contributes temperature dependent magnetic moment in $Eu^{3+}$ ions [12]. The alteration of ionic radii and valence states between $Eu^{2+}$ (~1.46 Å, 12-fold coordination) and $Eu^{3+}$ (~0.945 Å, 6-fold coordination) ions can allow a partial substitution of the Eu ions into the $Ba^{2+}$ (~1.61 Å, 12-fold coordination) sites or $Fe^{3+}$ (~0.645 Å, 6-fold coordination) sites [13, 14]. Various techniques, including sol–gel method [7, 15 - 17], hydrothermal synthesis [18], and solid-state reaction [19], have been employed to substitute the Eu ions to M-type hexaferrite structure.

Khademi et al. [15] observed the enhancement of coercivity from 1.92 kOe to 6.12 kOe, and decrease of magnetization from 59.21 emu/g to 45.14 emu/g with the increase of Eu atoms at the Ba sites of $BaFe_{12}O_{19}$, whereas Zafar et al. [18] and Liu et al. [19] has observed the enhancement both in magnetization and coercivity, and decrease in electrical conductivity by substituting Eu at the Ba sites. The co-doping of RE and transitional elements (Ni, Co) has been used for substituting at the Ba- and/or Fe-sites, and applying the materials for electro-magnetic and photo catalytic devices [20 - 23]. The RE atoms (La, Ce, Sm, Nd, etc) generally occupy at the Fe-sies, which results in a significant lattice distortion and strain and reduces magnetization in $BaFe_{12}O_{19}$ structure [24]. An excessive amount of RE atoms substitution produces secondary phases like $RFeO_3$ and α-$Fe_2O_3$ [25]. The non-stoichiometric atomic ratio at the Fe- and Ba-sites and lattice defects strongly control the magnetic exchange interactions, coupling between the FM and AFM phases, and electronic charge conduction mechanisms in the $BaFe_{12}O_{19}$ structure [26 - 28]. A controlled variation of the co-existing of α-$Fe_2O_3$ (AFM) and hexaferrite (FM) phases can effective by tune the magnetization, coercivity and exchange coupling of the Ba-hexaferrite material [8, 26, 29, 30, 31].

The present work has been designed to vary the atomic ratio of Eu and Ba in the $BaFe_{12}O_{19}$ structure. The first objective is to study the structural phase stabilization and determine the volume fractions of the M-type hexaferrite phase and the secondary (α-$Fe_2O_3$) phase. The second objective is to study the role of the secondary phase and non-stoichiometric atomic ratio on the control of electronic charge compensation, oxygen vacancy and lattice defects for tuning the magnetic super-exchange interactions, exchange coupling between FM and AFM units, magnetic moment, magnetic anisotropy, electronic charge transport, optical band gap, and magneto-electric properties.

## 2. Experimental

### 2.1. Sample preparation

The Eu substitution in $BaFe_{12}O_{19}$ structure was performed by using the sol-gel method [32, 33]. The starting precursors included barium nitrate $(Ba(NO_3)_2)$, europium nitrate hexahydrate $(Eu(NO_3)_3.6H_2O)$, and iron nitrate nonahydrate$(Fe(NO_3)_3.9H_2O)$. Both Ba-stoichiometric and non-stoichiometric compositions were prepared by following chemical reactions:

$$\left.\begin{aligned} x\,Ba(NO_3)_2 + y\,Eu(NO_3)_3.6H_2O + 12\,Fe(NO_3)_3.9H_2O \rightarrow\ Ba_xEu_yFe_{12}O_{19} \\ \text{Stoichiometric } (x,y) = \{(1,0),(0.9,0.1),(0.8,0.2)\} \\ \text{Non-stoichiometric } (x,y) = \{(1,0.2),(0.7,0.2),(0.6,0.2)\} \end{aligned}\right\} \quad (1)$$

For each composition, the appropriate amounts of nitrates were dissolved in distilled water. Chelating solution was prepared by mixing of citric acid and ethylene glycol in a 1:1 molar ratio into distilled water. Both solutions were separately stirred at 50 °C for 1 h to ensure homogeneity. The nitrate solution was further heated at 80 °C, and after 2 h, the chelating solution was added dropwise under continuous stirring. The mixture was heated at 120 °C for 4 h until a wet gel was formed, which subsequently dried into a greenish powder. The obtained powder was annealed at 200 °C to remove organics, during which the powder changed from green to black upon cooling. The black powders were manually ground for 1 h in an agate mortar, pelletized (7 mm diameter, ~0.5 mm thickness) under a force of 4 tons. The pellets were and subjected to multi-step annealing at 450 °C for 3 h, 950 °C for 4 h, and finally at 1200 °C for 24 h in air with a heating and cooling rate of 3 °C/min. The samples with stoichiometric compositions $BaFe_{12}O_{19}$, $Ba_{0.9}Eu_{0.1}Fe_{12}O_{19}$, and $Ba_{0.8}Eu_{0.2}Fe_{12}O_{19}$ were denoted by BM, B09E01, and B08E02, respectively. The samples with non-stoichiometric compositions $BaEu_{0.2}Fe_{12}O_{19}$, $Ba_{0.7}Eu_{0.2}Fe_{12}O_{19}$ and $Ba_{0.6}Eu_{0.2}Fe_{12}O_{19}$ were denoted by B10E02, B07E02, and B06E02, respectively.

2.2. Material characterization and measurements

X-ray diffraction (XRD) pattern was recorded by using a Rigaku X-ray diffractometer with Cu Kα source (λ = 1.5406 Å) in 2θ range of 20° to 80°. Micro-Raman spectra were recorded in the wavenumber range of 60 – 800 $cm^{-1}$ using a Raman microscope (Renishaw, UK) with a Laser source of wavelength 785 nm at 0.5 mW power with 30 s exposure time on the sample surface. The surface chemical properties were studied by X-ray photoelectron spectroscopy (XPS) using a model of K-Alpha-KAN9954133 (Thermo Scientific). The absorbance of UV-Visible spectra was recorded by SHIMADZU UV-Vis-NIR Spectrometer in the wavelength range 200-800 nm. DC magnetization measurements were performed using a Physical Property Measurement System (PPMS) (Evercool-II, Quantum Design, USA) in the temperature range of 3 - 350 K and magnetic fields up to ± 90 kOe. Temperature dependent electrical conductivity (σ) and current density vs. electric field (J-E) characteristics at selected temperatures were measured in the temperature range of 30 K to 300 K using Keithley 6517B electrometer. The silver paste was coated on both side of the pellets (7 mm diameter, ~0.5 mm thickness) for good electrical contact and sandwiched between top and bottom copper electrodes for electrical measurements. The low temperature was controlled using a closed cycle refrigerator (CCR) (Cryo Industries, USA) and measured using a Lakeshore temperature controller (model 336). Magnetoresistance (MR) measurements were performed using the Keithley 6517B while sample was placed between the poles of an electromagnet (Microsense, USA). The electrical and magneto-transport measurements were carried out using a custom designed (automated) Python-based program for precise control and synchronization of the experimental setup and data acquisition. The magnetoelectric (ME) coupling of the sample was investigated by measuring the induced voltage ($V_{out}$) across the sample using a Lock-in Amplifier (SR830, Standford Research). The sample was subjected to an external ac magnetic field ($H_{ac}$) generated by a Helmholtz coil pair, driven by the internal ac source of the Lock-in Amplifier. A superimposed dc magnetic field ($H_{dc}$) was applied using an electromagnet (MicroSense, USA). Both magnetic fields were applied along the in-plane direction of the sample, with surface of the pellet- shaped sample oriented perpendicular to the field direction, effectively minimizing inductance effects. The induced voltage ($V_{out}$) corresponding to the magnetoelectric response was recorded under dc magnetic field sweep, where the dc magnetic field ($H_{dc}$) was varied from 0 to 8 kOe at a fixed frequency and $H_{ac}$.

## 3. Results

### 3.1. X-ray diffraction pattern

Fig. 1 shows XRD patterns of the stoichiometric (BM, B09E01, and B08E02) and non-stoichiometric (B10E02, B07E02, and B06E02) samples. The undoped BM sample shows single-phased M-type hexaferrite structure. Rietveld refinement of the XRD patterns of the Eu substituted samples was performed by considering the M-type hexaferrite (magneto-plumbite structure with $P6_3/mmc$ space group) as the main phase, hematite (rhombohedral structure with space group R-3c) as the secondary phase, and $BaFeO_3$ (orthorhombic structure with space group Pbnm) as a minor phase [3, 34 - 36]. The $\chi^2$ values were found within 1.7 – 1.9. The information of secondary phase fraction is shown in the Fig. 1. The refinement of XRD patterns confirm that Eu substituted at Fe sites rather than Ba sites in $BaFe_{12}O_{19}$ structure. The chemical composition in the hexaferrite structure suggests excess amount of oxygen atoms (δ) in the range of 2 - 4 per formula unit. The non-stoichiometric amount of Fe and Ba atoms forms the secondary $\alpha$-$Fe_2O_3$ (hematite) phase (3.35 - 26.39 %) and $BaFeO_3$ (Ba-orthoferrite) (0.45 – 3.49 %), respectively, to optimize the m-type Ba-hexaferrite phase. Fig. 1(g-i) indicated a nearly linear increment of the total impurity phase with the increase of the Eu/Ba atomic ratio, except for the sample with excess Ba atoms (B10E02). The fractions of the co-existing impurity phase were reciprocating to optimize the M-type hexaferrite phase and adjust the local chemical structure. The structural parameters for the M-type hexaferrite and hematite phases have been summarized in Table 1. For the hexaferrite phase, the lattice parameter '*a*' decreased with the decrease of Ba-atomic concentration both in stoichiometric and non-stoichiometric compositions, whereas the lattice parameter along the *c*-axis increased in the stoichiometric compositions but decreased in the non-stoichiometric compositions. Overall, the unit cell volume of the m-type hexaferrite structure decreased with the increase of Eu/Ba atomic ratio. However, the c/a ration remains constant (~ 3.94) in the magneto-plumbite structure for all the samples [3, 32]. For the hematite phase, both lattice parameters (*a* and *c*) and the unit-cell volume decreased with the increase of hematite phase fraction both in stoichiometric and non-stoichiometric compositions of the material. The *c/a* ratio remains close to 2.73 in all samples.

Apart from cell parameters, the chemical structure of the metal-oxide (M-O) bond length, distortions and angle (∠M-O-M, ∠O-M-O) in five crystallographic $Fe^{3+}$ sites (2a (Fe1), 2b (Fe2), $4f_1$ (Fe3), $4f_2$ (Fe4), and 12k (Fe5)) are expected to be affected by the substitutional effect. In hexaferrite phase, the average length of Ba–O, Fe3-O, and Fe5-O bonds increased in the stoichiometric compositions and decreased in the non-stoichiometric compositions, whereas Fe1-O, Fe2-O, and Fe4-O monotonically decreased with the increase of Eu/Ba atomic ratio. The bond distribution, especially displacement of the oxygen atoms, is found to be relatively high for the 2b bipyramidal Fe-sites, 12k octahedral Fe-sites and the Ba-sites, in comparison to a minor distortion at the 2a octahedral Fe-sites,$4f_1$ tetrahedral and $4f_2$ octahedral Fe sites. A noticeable change has observed in the ∠Fe–O–Fe bond angles at different lattice sites of the hexaferrite structure has observed upon Eu substitution. The Fe–O–Fe bond angles below 120° favor the ferromagnetic (FM) super-exchange interaction, whereas the angles above 120° promote the antiferromagnetic (AFM) super-exchange interactions [34]. Accordingly, the Fe moments at the 12k sites (Fe5) prefer FM super-exchange interactions with Fe-moments at the 2a-,2b-, and 12k-sites, whereas AFM super-exchange prefers between the Fe-moments at the $4f_1$-2a, $4f_1$-12k sites, and 2b-$4f_2$ sites. The AFM super-exchange interaction between the 2a-$4f_1$ and 2b-$4f_2$ sites are slightly affected (diluted) by the increase of Eu/Ba ratio, whereas the super-exchange interactions between $4f_1$-12k sites are nearly unaffected. However, the presence of the canted AFM hematite phase plays a important role in modifying exchange interactions at the interfaces between two particles of M-type hexaferrite phase and magnetically ferrimagnetic spin order.

Lattice strain ($\varepsilon$) and Crystallite size (D) were determined by application of Size-Strain plots: $(\mathrm{d}\beta\cos\theta)^2 = \frac{K\lambda}{D}(\mathrm{d}^2\beta\cos\theta) + \frac{\varepsilon^2}{4}$ [37], using the peak profile parameters (position, $\beta$) from selected prominent peaks in the XRD patterns provided the crystallite size ($D$) in the range of 50 - 88 nm and 19 – 144 nm for the hexaferrite phase and hematite phase, respectively. On the other hand, the analysis shows the compressive lattice strain for both hexaferrite and hematite phases. It may affect on the contraction of cell volume for both the phases with the increase of Eu/Ba atomic ratio.

3.2. Raman spectra

The Raman spectrum in the pure M-type hexaferrite phase of the BM sample has been characterized by the 17 bands ($1E_{1u}$ + $5E_{1g}$ + $4E_{2g}$ + $7A_{1g}$) out of the theoretically predicted 42 bands ($11A_{1g}$+$14E_{1g}$+$17E_{2g}$) [38]. The observed bands in the BM sample are identified in Fig. 2. The 5 distinct bands of the hematite phase ($3E_g$ + 2 $A_{1g}$) involved in the motion of the Fe-O bonds within a single $Fe_2O_3$ unit [39, 40]. The undoped $BaFe_{12}O_{19}$ sample exhibited the vibrational modes designed for the hexaferrite structure, whereas Eu-doped samples displayed the vibrational modes for both hexaferrite and hematite structures. The bands in the hexaferrite phase and hematite phase of all the samples are summarized in Table 2. To understand the modifications in the chemical bonds in the hexaferrite phase and coexisting secondary phase ($\alpha$-$Fe_2O_3$), the profile parameters (peak position, full width at half maximum (FWHM ($\beta$)), and normalized intensity (normalized with respect to the highest intensity peak)) were extracted by deconvoluting the spectra of selective Raman-active modes. In the M-type Ba-hexaferrite lattice structure, the $A_{1g}$ (11 active) modes involve symmetric starching and atomic displacements along the c-axis (out-of-plane). The $E_{1g}$ (14 active) modes involve doubly degenerate in-plane vibrations of atoms. The $E_{2g}$ (17 active) modes doubly degenerate shearing or rational shift in the hexagonal planes. In the hematite phase, the $A_{1g}$ mode at ~ 227 $cm^{-1}$ and at 500 $cm^{-1}$ is attributed to the symmetric stretching vibrations of the Fe-O bonds along the c-axis. The $E_g$ modes at ~ 245 $cm^{-1}$ and ~ 293 $cm^{-1}$ correspond to in-plane vibrations of the Fe-O units. The $E_g$ mode at ~ 415 $cm^{-1}$ is associated with internal Fe–O bending vibrations within the octahedral units. To elucidate the influence of Eu substitution, the profile parameters were normalized with respect to the values of the hexaferrite phase in the undoped BM sample and their variations with the Eu/Ba ratio are plotted in Fig. 3 for some selected in-plane and out-of-plane Raman active modes of the hexaferrite phase. The in-plane vibrational modes exhibit pronounced changes with increasing Eu/Ba ratio. As shown in Fig. 3(a), peak 7 (~ 286 $cm^{-1}$) displays a gradual increase in the normalized Raman shift from unity to approximately 1.03, indicating a slight hardening of the vibrational mode. Simultaneously, its normalized intensity increases dramatically, reaching nearly eight times that of the undoped sample (Fig. 3(b)), while the normalized FWHM decreases continuously to nearly half of its original value (Fig. 3(c)). In contrast, peak 12 (~ 420 $cm^{-1}$) exhibits a progressive reduction in the Raman shift, accompanied by a moderate increase in the intensity and FWHM up to Eu/Ba ≈ 0.25, followed by a decrease in the FWHM at higher Eu concentrations. Peak 16 (~ 529 $cm^{-1}$) shows almost no variation in the Raman shift throughout the composition range. However, its intensity gradually decreases and the FWHM initially increases to about twice the value of the undoped sample before slightly decreasing at higher Eu/Ba ratio. The out-of-plane modes (peaks 9, 18, 19, and 20) exhibit comparatively smaller variations in Raman shift but substantial changes in intensity and FWHM. As shown in Fig. 3(d), the peaks 9 (~ 338 $cm^{-1}$), 19 (~ 685 $cm^{-1}$), and 20 (~ 718 $cm^{-1}$) undergo only minor frequency shifts (< 0.3%), whereas peak 18 (~ 615 $cm^{-1}$) exhibits a slight softening at higher Eu concentrations. The intensity variations in Fig. 3(e) reveal a continuous decrease for peaks 9, 19, and 20 with increasing Eu/Ba ratio, indicating a reduction in Raman scattering efficiency associated with these out-of-plane vibrations. In contrast, peak 18 exhibits a distinct behaviour, showing an abrupt enhancement in intensity for Eu/Ba ≥ 0.30 and reaching approximately 1.35 times the value of the undoped sample. This anomalous increase suggests a significant modification of the local environment of the $4f_2$ octahedral site. The FWHM evolution in Fig. 3(f) further supports these observations. Peaks 9 and 19 displayed a substantial broadening with increasing Eu/Ba ratio, indicating enhanced lattice distortion and reduced phonon lifetime. Peak 19 exhibits the largest broadening, reaching nearly three times the FWHM of the undoped sample before decreasing at higher

Eu/Ba ratio. Conversely, peaks 18 and 20 showed only marginal changes in FWHM, suggesting comparatively lower sensitivity of these modes to Eu-induced disorder. The in-plane modes are particularly sensitive to changes in structural ordering and bond strength, whereas the out-of-plane modes reflect the modifications of the octahedral framework along the c-axis. The anomalous intensity enhancement of peak 18, together with a significant broadening of peaks 9 and 19, indicates that Eu substitution introduces anisotropic lattice distortion and alters the local symmetry of the Fe–O polyhedral.

3.3. X-ray photoelectron spectroscopy

The survey scan (Fig. S1(a)) confirms the presence of expected elements Ba, Fe, Eu, and O in the samples. The charge correction was performed using the Carbon 1s peak at 284.80 eV. The O 1s spectra (Fig. S1(b) were deconvoluted into three distinct components. The low-binding-energy peak at ~529.56–529.84 eV is attributed to lattice oxygen ($O^{2-}$) involved in metal–oxygen bonding. The intermediate component appearing at ~ 530.67–531.40 eV is associated with oxygen ions in the vicinity of oxygen-vacancy ($V_O$) sites and surface defects. The high binding energy component at ~ 531.61–532.52 eV is assigned to surface hydroxyl groups and chemisorbed oxygen species [16, 41]. The relative peaks area (Table 3) was used to quantify the oxygen vacancy/surface defects using $V_O(\%) = \frac{A_{V_O}}{A_{V_O} + A_{M-O}} \times 100$, where $A_{M-O}$ and $A_{V_O}$ are area under metal oxide peak and oxygen vacancy peak, respectively. The oxygen deficiency decreases from 34 % at Eu/Ba = 0 to 21 % at Eu/Ba = 0.11 and reaches its lowest value of 9% at Eu/Ba = 0.20. With further increase in the Eu/Ba ratio, the relative oxygen vacancy increases up to 35% at Eu/Ba ratio 0.33.

The core levels Ba 3d, Eu 3d, and Fe 2p spectra in Fig. 4 were deconvoluted to obtain the profile parameters are summarized in Table 3. The characteristic 3d and 2p spectra showed $3d_{5/2}$, $3d_{3/2}$ and $2p_{3/2}$, $2p_{1/2}$ doublets, respectively, arising from the spin-orbit coupling. The binding energies (BE) of the Ba $3d_{5/2}$ and Ba $3d_{3/2}$ peaks are approximately 778.76 – 780.26 eV and 794.09 – 795.74 eV, respectively, and spin-orbit splitting energy (Δ) ~ 15.16 - 15.53 eV. These values are close to the expected BE of ~ 780 eV and ~ 795 eV for ionic $Ba^{2+}$ - $O^{2-}$ bonding, indicating $Ba^{+2}$ charge state [42]. The asymmetric shape of each Ba doublet resolved into two subcomponents, and the effect is more prominent for the sample (B06E02) with the highest Eu/Ba ratio. $Ba^{2+}$ is a nonmagnetic ion without unpaired valence electrons, and the sub-component in each doublet component confirm the co-existence of two distinct local crystal-field environments around the Ba ions and different Ba–O covalent bonding [3, 34]. The Eu 3d spectra are absent in BM and appear in all the Eu-substituted samples. The Eu 3d band splits into Eu $3d_{5/2}$ and Eu $3d_{3/2}$ components at binding energies ~ 1134.15 – 1135.04 eV and 1163.79 – 1164.56 eV, respectively. The spin–orbit splitting energy is in the range of 29.4 – 30.1 eV. These binding-energy values are characteristic of $Eu^{3+}$ ions in the trivalent oxidation state [19, 43]. The Fe 2p spectra exhibit a more complex structure of Fe $2p_{3/2}$ and Fe $2p_{1/2}$ doublets together with pronounced satellite peaks. The Fe $2p_{3/2}$ peak is located at ~ 710.25 – 714.77 eV, while the Fe $2p_{1/2}$ peak appears at 723.72 – 726.11 eV with spin–orbit splitting energy 13.0 - 13.7 eV. Satellite features are observed in the binding energy ranges of 718.49 – 718.99 eV and 730.55 – 733.12 eV for the Fe $2p_{3/2}$ and Fe $2p_{1/2}$ regions, respectively. The features of Fe 2p band are consistent with $Fe^{3+}$ ions in oxide environments [44 - 46]. Unlike the Ba 3d spectra, the Fe 2p spectra displayed multiplet splitting in both the Fe $2p_{3/2}$ and Fe $2p_{1/2}$ regions. These multiplet structures arise from the interaction of the Fe 2p core hole with unpaired 3d valence electrons and influenced by crystal-field effects, magnetic exchange interactions, and Fe-O covalent hybridization.

### 3.4. Optical properties

Fig. 5(a) shows the UV-visible absorbance spectra of the Eu substituted samples in the UV-Visible range of light. The undoped BM sample shows relatively strong absorption at lower wavelengths, followed by a gradual decrease in the absorbance at wavelengths higher than 400 nm. Compared with the undoped BM sample, the Eu-substituted samples show enhanced absorbance throughout the measured wavelength range, and characterized by a sharp absorption at wavelengths below 700 nm and the absorption feature at lower wavelength transform from a slow increasing nature to a nearly wavelength independent feature depending on the Eu/Ba ratio. Among the samples, B08E02 shows the highest absorbance, followed by B10E02, whereas BM shows the lowest absorbance. The enhanced absorption in the Eu-substituted samples suggests that Eu incorporation modifies the electronic structure and promotes additional optical transitions. The optical band gap ($E_g$) was estimated using Tauc's equation: $\alpha h\vartheta = A\left(h\nu - E_g\right)^n$, where α is the absorbance coefficient, A is the transition probability, h is the Planck's constant, $\nu$ is the frequency. A linear fit in the Tauc plots of $(\alpha h\nu)^{1/n}$ vs. $h\nu$ for n = ½ suggest direct electronic transition with optical band gap energy ($E_g$) (Fig. 5(b)). The direct band gap of the undoped BM sample is found to be 3.09 eV. Among the Eu substituted samples, the stoichiometric samples showed a decrease of the band gap from 3.09 eV to 2.25 eV with the increase of Eu content, whereas the non-stoichiometric samples showed an increase of band gap from 2.40 eV to 2.58 eV. This indicates that the modification of the electronic structure by Eu substitution is governed not only by the Eu/Ba ratio but also by structural deficiency.

The valence band spectra in the binding energy range of -10 eV to 20 eV were extracted from XPS survey scans. The valence band maximum ($E_{VBM}$) with respect to Fermi level ($E_F$), i.e., $E_F$-$E_{VBM}$, was determined by linear extrapolation of the leading edge of the valence band to the baseline, as shown in Fig. 5(c). The conduction band minimum ($E_{CBM}$) relative to the Fermi level were estimated using the optical bang gap ($E_g$) according to the relation: $E_{CBM} = E_g - (E_F - E_{VBM})$. The resulting band alignment diagrams are presented in Fig. 5(d). The work function ($\varphi$) of the samples was estimated using the relation: $\varphi = \chi + E_{CBM}$, where $\chi =$ 4.5 eV is electron affinity of $PbFe_{12}O_{19}$ [47]. The calculated work function ($\varphi$) values are 6.99 eV, 6.48 eV, 5.76 eV, 6.15 eV, 5.96 eV, and 6.45 eV for BM, B09E01, B08E02, B10E02, B07E02, and B06E02, respectively. The variation of the valence band maximum ($E_{VBM}$), conduction band minimum ($E_{CBM}$), and work function ($\varphi$) with the variation of the Eu/Ba ratio reveals a significant modification in the electronic band structure of the hexaferrite system. The ($E_F - E_{VBM}$) value increases from - 0.60 eV for BM (Eu/Ba = 0) to a maximum -1.10 eV at Eu/Ba = 0.29. However, at the highest Eu/Ba ratio of 0.33 (B06E02), the ($E_F - E_{VBM}$) value decreases to - 0.63 eV, suggesting a partial recovery towards the undoped BM value. The $E_{CBM}$ decreases from 2.49 eV (Eu/Ba = 0) to 1.26 eV (Eu/Ba = 0.25), followed by an increase at higher Eu/Ba ratio, suggesting a non-monotonic modification in the band structure. A similar trend is observed for the work function ($\varphi$), which decreases from 6.99 eV (Eu/Ba = 0) to 5.76 eV (Eu/Ba = 0.25), followed by a partial recovery to 6.45 eV for B06E02 (Eu/Ba = 0.33). Despite these variations, the Fermi level remains closer to the valence band than to the conduction band for all samples, confirming the persistence of p-type semiconducting behaviour in the Eu-substituted samples.

### 3.5. Magnetic properties

Fig. 6 shows the temperature (T) dependence of the zero-field cooled (ZFC) and field cooled (FC) magnetization (M) curves of the samples at applied magnetic field of 250 Oe. In the ZFC mode, the samples were cooled from 300 K to 3 K without an external magnetic field, whereas in the FC mode the samples were cooled from 350 K to 3 K under a constant magnetic field (250 Oe). The samples exhibited thermomagnetic irreversibility between the FC and ZFC curves below the maximum measurement temperature of 350 K. The irreversibility effect increases on decreasing the temperature, where the ZFC curves exhibited a peak-like behaviour at temperature $T_m$ below which ZFC magnetization decreases due to spin freezing and anisotropy effect and the FC magnetization either decreases or levels off at temperature below $T_m$ [19]. The position of the broad maximum is shifts from 58 K to 43 K with the increase of Eu/Ba ratio in the hexaferrite phase, except in B07E02 sample containing Eu/Ba = 0.29 shows at 55 K. The thermomagnetic irreversibility effect is quantified by the subtraction of $M_{ZFC}$ from the $M_{FC}$ curves, i.e., $\Delta M = M_{FC} - M_{ZFC}$. The temperature variation of the $\Delta M$ in the insets of Fig. 6 shows the positive remanent magnetization value throughout the measurement temperatures up to 350 K and the value decreases with the increase of temperature. The nature of the variation of the $\Delta M(T)$ curves, depends on the Eu/Ba ratio, especially a low temperature upturn suggests a strong spin freezing/pinning effect in the samples. The undoped BM sample exhibits the largest $\Delta M$ (~ 0.64 emu/g) at 3 K, indicating strong domain-wall pinning and magnetic anisotropy effect. Upon Eu substitution, the $\Delta M$ value at 3 K decreases into the range of 0.08 – 0.5 emu/g, suggesting a reduction in the magnetic spin pinning effect. On other hand, the reduction of the bulk $M_{ZFC}$ or $M_{FC}$ values at 3 K in the Eu substituted samples can be affected by the increase of hematite phase, which is magnetically in the AFM or weak FM phase [48].

The effects of Eu substitution and the secondary hematite phase in the hexaferrite structure can be realized by saturating the bulk magnetization ($M_{sat}$) and other ferrimagnetic parameters (coercivity ($H_C$), remanent magnetization ($M_R$), squareness ratio ($M_R/M_{sat}$), loop area ($\int M\, dH$)) of the samples at higher magnetic fields. The magnetic field dependence of magnetization (M(H)) curves at 3 K and 300 K (Fig. 7) exhibited well defined hysteresis loops and typical characteristics of ferrimagnetic (FM) order in the M-type $BaFe_{12}O_{19}$ structure [19]. The magnetization approaches to saturation at magnetic fields (Hs) typically above ~ 20 – 30 kOe. The saturation magnetization ($M_{sat}$) was calculated by using the law of approach to saturation of magnetization, $M(H) = M_{sat}(1- b/H^2)$ for H > 20 kOe, where the parameter b = $\frac{8K_{eff}^2}{105M_{sat}^2}$ provides the effective anisotropy constant ($K_{eff}$) [3, 34]. The ferrimagnetic parameters, as shown in Table 4, exhibited a pronounced dependence either on the Eu/Ba ratio or hematite phase. A systematic decrease of the $M_{sat}$ values is consistent with the increase of the secondary hematite phase fraction, rather the increase of Eu/Ba ratio [32]. However, rest of the ferrimagnetic parameters are controlled by the non-stoichiometric Eu substitution in the hexaferrite structure. The chemical compositions of excess Eu content exhibited the highest value of the ferrimagnetic parameters, whereas the chemical composition with smaller amount of Eu content showed the smaller values of the ferrimagnetic parameters. The variation of the saturation magnetization follows the composition rule $M_{sat} = f_{FM}M_{FM} + (1 - f_{FM})M_{AFM}$, where $M_{AFM}$ is saturation magnetization of AFM hematite phase ($M_R$ ~ 0 emu/g at 3 K, and $M_R$ ~ 0.17 emu/g at 300K [49]), $f_{FM}$ and $M_{FM}$ is volume fraction and saturation magnetization of the ferrimagnetic pure BM sample, respectively [19, 31, 50]. The squareness ratio ($M_R/M_{sat}$) remains nearly same both at 3 K and 300 K (0.25 – 0.50). The low squareness ratio (~ 0.25 – 0.34) in the pure BM sample and the highest Eu -deficient sample indicates multi-domine behaviour, whereas the squareness ratio (~ 0.46 – 0.50) satisfied the Stoner-Wohlfarth limit ($M_R/M_{sat}$ ~ 0.46 - 0.50), for weakly interacting single domain character of the ferrimagnetic domains [9, 31, 51], which fully utilizes the intrinsic magneto-crystalline anisotropy for domine reversal mechanism [30, 31, 50]. The $M_{sat}$ value of the samples decreased at 300 K, but the coercivity ($H_C$) and anisotropy constant ($K_{eff}$) increased at 300 K. This results in the increase of the energy product (loop) area of the samples at 300 K with reference to 3 K.

3.6. Electronic charge conduction properties

3.6.1. Temperature dependence of electrical conductions.

The temperature dependence of dc conductivity (σ) was investigated by measuring the current (I) under an applied electric field of 20 kV $m^{-1}$ (10 V) during heating of the samples from 30 K to 300 K at a ramp rate of 1 K $min^{-1}$. The dc conductivity was calculated using: $\sigma =$

$\frac{\mathrm{I} \times \mathrm{t}}{V \times A}$, where ‘t’ is the sample thickness, ‘V’ is the applied voltage, and ‘A’ is the cross-sectional area of the sample. The conductivity in Fig. 8 showed a non-linear increase with temperature. This confirms a typical insulators or semiconductor behavior of the samples. The low temperature feature is dominated by nearly temperature independent Insulator (I) character (Region-I), which transforms into temperature-dependent semiconductor (SM) character (Region-II) and marked by an insulator to semiconductor transition temperature ($T_{IS}$). The $T_{IS}$ values were determined from the activation energy $\left(\Delta \mathrm{E} = -\frac{\mathrm{d}(\mathrm{Ln}[\sigma])}{\mathrm{d}\left(\frac{1}{K_B T}\right)}\right)$ curves (Fig. 8), where the activation energy makes a sharp jump from nearly zero level to high energy. As shown in the inset of Fig. 8(a), the $T_{IS}$ values decreased with the increase of the Eu/Ba ratio, except an unusual increase for the non-stoichiometric sample B10E02 and a decrease for the non-stoichiometric B07E02 sample. Based on the temperature dependence of the local activation energy ΔE(T), the transport regions of the BM sample exhibits three regions: Region-I (30–133 K) with ΔE ~ 1 – 20 meV; Region-II (133–233 K) with ΔE ~ 0.02 - 0.40 eV; and Region-III (> 233 K) with ΔE > 0.40 eV and a peak at T ~ 265 K with ΔE ~ 1.30 eV. The Eu-substituted samples exhibited only the low temperature insulator region (Region-I) and the high temperature semiconductor region (Region-II) with a gradual increase of activation energy (ΔE). The B10E02 sample showed a peak of ΔE ~ 1.40 eV at the same temperature ~ 265 K as noted for the BM sample. The conductivity of the samples in the insulator regime and semiconductor regime lies in the range of ~ $10^{-12}$ – $10^{-9}$ S/m, and ~ $10^{-8}$ – $10^{-4}$ S/m, respectively. The charge conduction mechanism in the temperature scale was understood by analyzing σ(T) data using the equation [41, 52]:

$$\sigma(T) = \sigma_o T^q \exp\left(-\left[\frac{T_o}{T}\right]^p\right) \tag{2}$$

where the parameters q and p define the nature of distinct conduction processes such as long-range charge hopping (q = 0, p = 1), Small polaron hopping (SPH) (q = -1, p = 1), Mott variable range hopping (MVRH) (q = -1/2, p = ¼,) and Efros-Shklovskii variable range hopping (ESVRH) (q = -1, p = ½). In the SPH model, strong carrier–lattice interactions localize charge carriers, and transport occurs via thermally activated hopping between neighbouring sites. In MVRH, charge carriers hop between localized states over variable distances in a disordered system with a nearly constant density of states near the Fermi level [53]. The ESVRH extends the MVRH model by incorporating Coulomb interactions among the localized charge carriers, which create a Coulomb gap near the Fermi level. We have observed that the low temperature region (~ 37–120 K) is well described by the ESVRH model (Fig. 9(a-f)). As the temperature increases, the conductivity follows the MVRH model in the intermediate temperature range (~ 130–300 K) (Fig. 9(g-l)). In contrast, high-temperature conduction region (~250–300 K) of the BM and B10E02 samples is well fitted by SPH model (Fig. S2). This transition indicates that, at sufficiently high temperatures, charge carriers acquire enough thermal energy for assisting in the polar hopping. Although three conduction mechanisms (ESVRH, MVRH, and SPH) are identified in B10E02 sample from the fitting of $\sigma(T)$ data, the activation energy (ΔE) analysis revealed only two transport regions. This discrepancy can be attributed to the comparable characteristic temperatures ($T_o \sim 10^3 - 10^4$ K) obtained for the ESVRH and MVRH models. The absence of SPH behaviour in the samples with higher hematite phase content samples suggests that increased defect density and disorder suppress the coherent polaron motion, and follow the MVRH dominated transport mechanism even at higher temperatures. The characteristic Mott temperature ($T_o$) was used to estimate the density of localized states near to Fermi level ($N(E_F)$), hopping distance ($R_H$), and hopping energy ($W_H$) using the equations [41]: $T_o = \left[\frac{24}{k_B N(E_F)\xi^3}\right]$, $R_H = \left[\frac{\xi}{8\pi k_B T N(E_F)}\right]^{1/4}$, $W_H = 0.25 k_B T_o^{1/4} T^{3/4}$, respectively. Here, $\xi$ is the localization length, which was taken as twice the average Fe-O bond length. The calculated

Mott's parameters at T = 200 K are summarized in Table 5. The condition for MVRH mechanism, $R_H > \xi$, was satisfies for all the samples, except B10E02. The samples have shown relatively similar values of the parameters with $N(E_F) \sim 10^{24}$ $eV^{-1}m^{-3}$, $R_H \sim 4.87 - 6.29$ nm, and $W_H \sim 0.26 - 0.34$ eV and consistent with MVRH model. However, the B10E02 sample deviates from the other compositions. It exhibits lower $T_0 \sim 10^4$ K with high density of localized states $N(E_F) \sim 10^{29}$ $eV^{-1}m^{-3}$. Its hopping distance is also considerably reduced to 0.30 nm, while the hopping energy decreases to 0.02 eV. These parameters indicate that the charge-transport mechanism in B10E02 differs substantially from that of the other Eu-doped samples and suggest that the MVRH model is not applicable to this composition.

3.6.2. Electric field dependence of electrical conductions.

The current density vs. electric field (J-E) characteristics were measured at selected temperatures using a triangular electric field waveform with the sequence 1: $(0 \rightarrow +E_{max}) \rightarrow$ 2: $(+E_{max} \rightarrow 0) \rightarrow$ 3: $(0 \rightarrow -E_{max}) \rightarrow$ 4: $(-E_{max} \rightarrow 0)$ at a sweep rate of 2 kV $m^{-1}$ $s^{-1}$. The J-E characteristics in the semi logarithmic plots (Fig. 10) exhibited a typical non-linear increase, confirming a semiconductor behaviour in the samples. A nearly symmetric J-E loops in the positive and negative sides of the fields indicate an identical electronic configuration at the interface between electrodes and both sides of the sample surface. The samples showed two distinct characteristics i.e., (1) resistive switching (hysteresis) and (2) resistive non-switching (non-hysteresis) behaviour in the J-E curves depending on the measuring temperature. In the resistive switching (RS), the J-E curves showed irreversible paths, consisting of a low resistance state (LRS) and high resistance state (HRS). The samples showed two types of resistive switching, (1) memristor switching and (2) polarization switching. In the first case, a high resistance state (HRS) during paths 1 and 3 are followed by the low resistance state (LRS) during reversing paths 2 and 4, respectively. The memristive switching [54, 55] can be attributed to inductive response from a trap mediated charge conduction process. In the case of polarization switching, the return paths follow the HRS and the behaviour is attributed to capacitive effect arising from the polarization current due to displacement of ions at the lattice positions and space charge polarization [54]. In the resistive switching behaviour, The J-E characteristics exhibited butterfly shaped loops with the current density minimum observed either on the J-axis (symmetric loop) or off centring from the J-axis (asymmetric loop). Depending on the transformation between resistive switching to non-resistive switching behaviour, the samples can be classified into groups. In one class (samples BM, B10E02 and B09E01), the non – resistive switching behaviour (symmetric curves without loop) was observed at lower temperatures (typically 100 K). It transformed into symmetric or asymmetric resistive switching behaviour at high temperatures (typically 200 K and above). In another class (samples B08E02, B07E02, B06E02), the (symmetric or asymmetric) resistive switching behaviour at the lower temperature (typically 100 K or lower) transformed into non-resistive switching behaviour at higher temperature (including 300 K). It is noted that the capacitive effect dominates at the RS behaviour at high temperatures and inductive effect dominates at the RS behaviour at lower temperatures. Overall, the evolution of the J–E loops from inductive to capacitive nature or vice-versa reflects a competition between charge trapping assisted conductivity with displacement-current-driven charge transport, space charge polarization effect, and thermally activated charge transport mechanisms in the material [56].

In order to understand the charge trapping and de-trapping assisted electrical conduction and electrode/sample interface effects, the J(E) curves at selected temperature (100 K, 200 K, and 300 K) were understood using the bulk-limited conduction (BLC) mechanisms, which include space charge limited conduction (SCLC) and Poole-Frenkel (P-F) emission, and the electrode-limited conduction (ELC) mechanisms, which include Schottky emission (SE) [56, 57]. The bulk-limited conduction mechanisms were understood by applying the space-charge-limited conduction (SCLC) mechanism, which follows the power law $J \propto E^m$ and a linear fit in the log

J vs. log E plots (Fig. 11). For all samples, the extracted $m$-values lie between 1 and 2, where m ~ 1 represents a typical ohmic conduction and the m ~ 2 represents the trap free SCLC mechanism. In contrast, the J-E curves of the B08E02, B07E02, B06E02 samples at 100 K during reverse path of the loops exhibited m values below 1. It indicates that charge transport in the low temperature regime can be affected by defect assisted Poole–Frenkel (P-F) emission or Schottky emission mechanisms. The P-F emission satisfies the equation [56, 57]:

$$J = qn_c\mu E\ exp\left(\frac{-q\varphi_T}{K_BT} + \frac{q\sqrt{\frac{qE}{\pi\varepsilon_O\varepsilon_r}}}{rk_BT}\right) \tag{3}$$

where $n_c$ is the density of charge carriers in the conduction band, T is the temperature, q is the charge of electron, $\varphi_T$ is the trap barrier potential, $\mu$ is charge carrier mobility, r is the coefficient (range:1–2) and $\varepsilon_r$ is the relative dielectric permittivity of material. The validity of P-F emission has been examined by linearity of ln(J/E) vs. $\sqrt{E}$ plot (Fig. S3). The P-F mechanism is mostly applicable at higher temperatures (200 K and 300 K) and in the forward path (1) of the J-E curves at higher electric fields (> 8 kV/m) and not applicable during reverse sweeps. The electrode limited Schottky emission mechanism is described by:

$$J = AT^2 exp\left[\frac{-q\left(\varphi_B - \sqrt{\frac{qE}{4\pi\varepsilon_O\varepsilon_r}}\right)}{k_BT}\right] \tag{4}$$

where A is Richerson constant (~ $8 \times 10^4$ A $m^{-2}$ $K^{-2}$ for metal oxide structure), $\varphi_B$ is the potential barrier formed at the interface of electrode and sample. A linear fit in the ln J vs. $\sqrt{E}$ plots (Fig. 12) suggest the application of SE mechanism for the present samples at higher electric fields (> 10 kV/m). The large work function (φ) mismatch between Ag electrodes (φ = ~ 4.26 eV) and samples (φ = 6.99 eV (BM), 6.48 eV (B09E10), 5.76 eV (B08E02), 6.15 eV (B10E02), 5.96 eV (B07E02), and 6.45 eV (B06E02)) resulted in Schottky barrier formation and exhibited non-linear increase of current density at higher electric fields. The conductivity (σ = J/E) at 20 kV/m, barrier height energy ($\varphi_B, \varphi_T$) at zero bias electric field and relative permittivity ($\varepsilon_r$) obtained from analysis of the J(E) curves at 300 K have been compared in Table 5. At 300 K, the electrical conductivity showed a systematic increase from $4.74 \times 10^{-6}$ S/m for Eu/Ba ratio = 0 to $4.33 \times 10^{-5}$ S/m for Eu/Ba ratio = 0.33. The Schottky barrier height ($\varphi_B$) values (0.65 – 0.73 eV) are found nearly twice of the trap barrier height ($\varphi_T$) ~ 0.28 - 0.34 eV obtained from the P-F mechanism. It means the material contains a high concentration of the defects and trapping centres that capture the moving electrons which are excited to the conduction band during sweeping of the electric field. On the other hand, the relative values of the dynamic permittivity ($1.50 \times 10^{-3}$ – $9.67 \times 10^{-3}$) obtained from SE mechanism are found to be nearly 100 times smaller than the values (0.29 – 1.57) obtained from the analysis of P-F mechanism. Such a large discrepancy suggests that P-F mechanism is not a suitable mechanism to explain the J-E characteristic at higher electric fields.

The XPS results confirmed that most Fe ions exist in the $Fe^{3+}$ state but fractional charge transfer between neighbouring Fe sites through $Fe^{(3-\delta)+}$ - $O^{2-}$ - $Fe^{(3+\delta)+}$ super-exchange pathways enhance the covalent nature of the Fe–O bonds. Furthermore, XPS analysis revealed the presence of oxygen vacancies, which act as donor-type defects and generate localized electronic states within the band gap. The formation of oxygen vacancies can be represented by using Kröger–Vink notation as: $O_0^{\times} \rightarrow V_0 + \frac{1}{2}O_2$, which can release free electrons through ionization process ($V_0 \rightarrow V_0' + e^-$ and $V_0' \rightarrow V_0'' + e^-$) [52, 58]. These electrons may participate in charge transport either through direct interaction with Fe ions or polaron formation. Consequently, oxygen vacancies serve as electron donors and create trapping centres that strongly influence the conduction process. Overall, the electrical transport in Eu-doped $BaFe_{12}O_{19}$ is primarily governed by the Schottky emission process at the electrode/sample interface and bulk conduction is affected by Poole–Frenkel emission through defect-related trap states.

3.6. Magneto-electric properties

The magneto-resistance (MR) measurements at room temperature were carried out by applying a constant voltage (~2 V) and measuring the corresponding current, while sweeping the dc magnetic field. The magneto-resistance ratio (MR%) was calculated using: $MR\ (\%) = \frac{\frac{V_{applied}}{I\ (H\neq0)} - \frac{V_{applied}}{I\ (H=0)}}{\frac{V_{applied}}{I\ (H=0)}} \times 100$, where $I\ (H \neq 0)$ and $I\ (H = 0)$ denote the measured current in the presence and absence of magnetic field, respectively, and shown in Fig. 13. The magneto-resistance was measured in two geometry modes such as, longitudinal mode (E || H) and transvers mode (E ⊥ H) as shown in inset of Fig. 13(d). The BM sample shows a non-monotonic field dependence in both geometries, with MR initially increasing and then decreasing at higher fields. In the Eu-doped samples, MR decreases monotonically with increasing field, except for B08E02, B10E02, which also shows crossover behavior. This crossover behavior was analyzed using the semi-empirical expression: $MR\ (\%) = \alpha H^2 + \frac{\beta H^{\gamma}}{1+\left(\frac{H_o}{H}\right)^n}$, where $\alpha$, $\beta$, $\gamma$ are fitting constants, $H_o$ represents crossover magnetic field, n is sharpness of crossover [59, 60]. Crossover fields were extracted for BM ($H_o$ = 1124 Oe for E || H, 664 Oe for E ⊥ H), B08E02 ($H_o$ = 2803 Oe for E || H), and B10E02 ($H_o$ = 1306 Oe for E || H) as marked in Fig. 13. These crossovers mark a transition from a low field regime dominated by classical, Lorentz force induced quadratic MR to a high field regime governed by spin polarized transport and magnetic saturation effects. The MR values at 8 kOe and fitting parameters ($\alpha$, $\beta$, $\gamma$) are shown in Table 6. In the longitudinal mode, all samples exhibit negative magnetoresistance, with the largest magnitude observed for the B10E02 sample (-51.36 %), followed by the BM (-37.59 %) and B06E02 (- 26.11 %), due to suppression of spin-disorder scattering and alignment of localized magnetic moments with increasing magnetic field [61]. In transverse mode BM and B10E02 show positive MR of 3.13 % and 33.12 %, respectively, while remaining Eu -doped samples show negative MR. The magnitude of MR in the longitudinal mode is larger than that in the transverse mode for all samples. This anisotropic MR behavior can be attributed to the difference in charge-carrier scattering mechanisms under different field geometries. In the E ∥ H geometry, enhanced spin-dependent scattering and orbital contributions associated with Fe 3d states lead to a larger scattering cross-section, resulting in stronger field-dependent resistance changes compared to the E ⊥ H geometry [62]. The fitting parameters reveal that the exponent $\gamma$ is positive for most samples in both geometries, indicating a super-linear contribution of magnetic field to MR at higher fields. Variations in the coefficients $\alpha$ and $\beta$ reflect changes in microstructural disorder, carrier localization, and magnetic anisotropy induced by Eu substitution and hematite phase formation. Overall, the observed anisotropic MR, crossover behaviour, and systematic evolution of fitting parameters confirm the strong coupling between charge transport and magnetic ordering in Eu-doped $BaFe_{12}O_{19}$.

Finally, magneto-electric (ME) coupling was studied at the room temperature by measuring of induced voltage across the sample while sweeping the dc magnetic field ($H_{dc}$) from 0 kOe to 8 kOe. The measurement performed at a fixed ac field amplitude of $H_{ac}$ = 4 Oe at frequencies of 973 Hz, 1973 Hz, and 3173 Hz. The magneto-electric coupling constant was calculated using: $\alpha = \frac{1}{t}\frac{V_{out}}{H_{ac}}$ [52], where t is the thickness of the sample. The normalized ME coupling constant $\left(\frac{\alpha}{\alpha_{min}}\right)$ versus dc magnetic field were shown in Fig. 14. The ME coupling constant exhibits a peak at a characteristic $H_{dc}$ field, corresponding to the field at which the ME coupling is maximum [52, 63]. For the undoped BM sample, the peak is observed at $H_{dc}$ ~ 2220 Oe at 973 Hz. With increasing Eu/Ba ratio, the peak position shifts towards higher magnetic fields, occurring at 3176 Oe for B09E01 (Eu/Ba = 0.11), 4500 Oe for B10E02 (Eu/Ba = 0.2), 5575 Oe for B08E02 (Eu/Ba =0.25), and 5770 Oe for B07E02 (Eu/Ba = 0.29). However, the

peak shifts back to a lower field of 2880 Oe for B06E02 (Eu/Ba = 0.33). The magnitude of the normalized peak increases with Eu/Ba ratio, from 1.06 for BM to more than 1.10 – 1.14 for Eu-doped samples, indicating enhancement in the field tunability of the ME coupling [63]. In addition, the $\alpha_{min}$ values (Fig. 14) increase with both Eu/Ba ratio and frequency. The B10E02 shows the highest $\alpha_{min}$ values of 8.38, 12.48, and 15.82 mV cm$^{-1}$ Oe$^{-1}$ at 973, 1973, and 3173 Hz, respectively. Unlike the sharp single resonance observed for BM and B09E01, the Eu-doped samples shows broader and multi-peaks around the peak region, which may be associated with the presence of multi-domain magneto-strictive response [64].

## Conclusions

The lattice structure, magnetic spin order, optical band gap and electronic charge conduction mechanism in the M-type $BaFe_{12}O_{19}$ hexaferrite structure has been modified by substituting stoichiometric and non-stoichiometric Eu atoms at the Ba-sites and by controlling the secondary hematite phase. The Rietveld refinement of XRD patterns and micro-Raman spectroscopy confirmed that the $Eu^{3+}$ ions preferentially occupied the $Fe^{3+}$ sites rather than the $Ba^{2+}$ sites. This substitution promoted the formation of secondary $Fe_2O_3$ (hematite) phase. X-ray photoelectron spectroscopy (XPS) confirmed the trivalent charge states in Fe and Eu ions, while the surface defects and oxygen vacancy concentration non-monotonically varied with the Eu/Ba ratio. The presence of antiferromagnetic hematite phase remarkably modified the ferrimagnetic properties of the M-type $BaFe_{12}O_{19}$ hexaferrite by exhibiting a soft to hard magnetic phase with the variation of Eu/Ba ratio. The dc electrical conductivity demonstrated a transition from insulator state to semiconductor state typically above 93 – 230 K, depending on the Eu/ Ba ratio. Detailed analysis revealed a crossover in the conduction mechanism from Efros-Shklovskii variable-range hopping (ESVRH) at low temperatures to Mott variable-range hopping (MVRH) at high temperatures. The non-linear J-E characteristics are driven by the electrode-limited Schottky emission at the interface and bulk-limited space charge effects and defect trap assisted conduction mechanism. The J-E hysteresis loops and resistive switching behaviors undergo a transition from inductive (memristive) to capacitive polarization characteristics depending on the measuring temperature and Eu/Ba ratio in the material. This study highlights that controlling of stoichiometric and non-stoichiometric rare earth atomic substitution and control of secondary ($\alpha$-$Fe_2O_3$) phase offer an effective strategy to tune the properties of hexaferrites and opens promising pathways for designing high-performance materials for spintronics, magneto-electronics, and multifunctional applications.

Acknowledgement

The authors acknowledge the research grant (NO: CSR-IC-ISUM- 03/CRS-286/2019-20/1338) from UGC-DAE CSR Indore, research grant (NO. CRG/2019/000505) from DST-SERB, Government of India for purchasing the low temperature cryo-stat, measurement facilities from CIF and Department of Physics at Pondicherry University, and UGC-DAE CSR Mumbai, for carrying out this work. GK acknowledges financial support from UGC-CSIR Fellowship.

Table. 1 Structural parameter of hexaferrite and hematite phases from Rietveld refinement of Eu doped Barium hexaferrite XRD patterns.

| Sample | BM | B09E01 | B08E02 | B10E02 | B07E02 | B06E02 |
|---|---|---|---|---|---|---|
| Hexaferrite phase (%) | 100.00 | 89.88 | 82.84 | 93.16 | 78.32 | 73.16 |
| Chemical composition | $BaFe_{12.11}O_{21.22}$ | $BaFe_{11.87}Eu_{0.12}O_{21.74}$ | $BaFe_{11.81}Eu_{0.20}O_{20.98}$ | $BaFe_{11.81}Eu_{0.19}O_{20.87}$ | $BaFe_{11.82}Eu_{0.20}O_{20.92}$ | $BaFe_{11.76}Eu_{0.25}O_{23.06}$ |
| Lattice parameter a (Å) | 5.89457(11) | 5.89245(9) | 5.89240(9) | 5.89524 (9) | 5.89163(11) | 5.88965(31) |
| Lattice parameter c (Å) | 23.21937(62) | 23.22821(50) | 23.22056(52) | 23.22229(48) | 23.21867(61) | 23.21419(184) |
| c/a ratio | 3.94 | 3.94 | 3.94 | 3.94 | 3.94 | 3.94 |
| Unit cell volume (Å$^3$) | 698.690(26) | 698.456(21) | 698.213(22) | 698.937(20) | 697.974(26) | 697.370(76) |
| Avg. $d_{Ba-O}$ (Å) | 2.7905(92) | 2.8079(2) | 2.8075(3) | 2.8084(2) | 2.8072(2) | 2.8065(2) |
| Avg. $d_{Fe1-O}$ (Å) (2a -site) | 1.9710(102) | 1.9666(2) | 1.9664(4) | 1.9670(3) | 1.9662(2) | 1.9656(4) |
| Avg. $d_{Fe2-O}$ (Å) (2b – site) | 2.0086(193) | 1.9989(5) | 1.9986(3) | 1.9992(5) | 1.9984(4) | 1.9979(4) |
| Avg. $d_{Fe3-O}$ (Å) (4f$_1$ - site) | 1.9077(154) | 1.9185(4) | 1.9183(5) | 1.9190(5) | 1.9181(5) | 1.9175(4) |
| Avg. $d_{Fe4-O}$ (Å) (4f$_2$ – site) | 2.0720(161) | 2.0577(6) | 2.0575(7) | 2.0583(6) | 2.0572(5) | 2.0566(7) |
| Avg. $d_{Fe5-O}$ (Å) (12k -site) | 2.0050(249) | 2.0056(2) | 2.0054(4) | 2.0061(6) | 2.0052(7) | 2.0046(3) |
| Ba-site distortion (× $10^{-4}$) | 16.633 | 12.649 | 12.727 | 12.821 | 12.716 | 12.685 |
| 2a(Fe1)-site distortion (× $10^{-4}$) | 0.001 | 0.000 | 0.001 | 0.000 | 0.001 | 0.001 |
| 2b(Fe2)-site distortion (× $10^{-4}$) | 117.705 | 128.915 | 128.597 | 128.198 | 128.644 | 128.769 |
| 4f$_1$(Fe3)-site distortion (× $10^{-4}$) | 0.203 | 0.079 | 0.087 | 0.097 | 0.086 | 0.083 |
| 4f$_2$(Fe4)-site distortion (× $10^{-4}$) | 0.203 | 0.782 | 0.772 | 0.763 | 0.776 | 0.776 |
| 12k(Fe5)-site distortion (× $10^{-4}$) | 34.304 | 25.952 | 25.942 | 25.929 | 25.942 | 25.948 |
| ∠ Fe2 – O1 – Fe5 (deg) | 121.60(18) | 121.30(5) | 121.29(6) | 121.28(7) | 121.29(8) | 121.30(2) |
| ∠ Fe3 – O2 – Fe5 (deg) | 126.20(14) | 126.60(3) | 126.59(4) | 126.58(3) | 126.59(5) | 126.59(9) |
| ∠ Fe5 – O2 – Fe5 (deg) | 89.00(3) | 88.09(2) | 88.10(4) | 88.12(1) | 88.10(2) | 88.09(6) |
| ∠ Fe2 – O3 – Fe4 (deg) | 145.30(16) | 144.93(5) | 144.94(2) | 144.95(1) | 144.94(1) | 144.93(8) |
| ∠ Fe4 – O3 – Fe4 (deg) | 82.70(12) | 82.28(3) | 82.26(5) | 82.24(1) | 82.26(7) | 82.27(5) |
| ∠ Fe1 – O4 – Fe3 (deg) | 126.50(10) | 125.68(9) | 125.70(3) | 125.72(1) | 125.70(1) | 125.69(5) |
| ∠ Fe1 – O4 – Fe5 (deg) | 96.00(3) | 96.75(9) | 96.74(7) | 96.73(1) | 96.74(9) | 96.75(4) |
| ∠ Fe5 – O5 – Fe5 (deg) | 98.00(7) | 98.20(5) | 98.21(7) | 98.23(3) | 98.21(5) | 98.21(1) |
| ∠ Fe4 – O5 – Fe5 (deg) | 119.00(4) | 120.62(2) | 120.61(7) | 120.61(1) | 120.61(8) | 120.62(2) |
| Lattice strain (%) | -0.31 | -0.17 | -0.24 | -0.23 | -0.28 | -0.30 |
| Crystallite size (nm) | 79 | 88 | 85 | 77 | 79 | 50 |
| Hematite Phase (%) | 0.00 | 8.86 | 13.90 | 3.35 | 20.07 | 26.39 |
| Lattice parameter a (Å) | -- | 5.03932 (21) | 5.03764(14) | 5.03986(77) | 5.03746(12) | 5.03690(40) |
| Lattice parameter c (Å) | -- | 13.75601(79) | 13.75093(60) | 13.76127(89) | 13.75121(51) | 13.75088(66) |
| c/a ratio | | 2.73 | 2.73 | 2.73 | 2.73 | 2.73 |
| Unit cell volume (Å$^3$) | -- | 302.529(25) | 302.215(18) | 302.710(38) | 302.201(15) | 302.126(50) |
| Avg. $d_{Fe-O}$ (Å) | -- | 2.0338(2) | 2.0331(1) | 2.0362(1) | 2.0331(1) | 2.0330(2) |
| ∠ Fe – O – Fe (deg) | -- | 85.60(4) | 85.60(1) | 85.79(1) | 85.60(5) | 85.61(4) |
| ∠ Fe – O – Fe (deg) | -- | 131.12(9) | 131.12(8) | 131.16(7) | 131.12(9) | 131.13(1) |
| Lattice strain (%) | -- | -0.20 | -0.18 | 1.59 | -0.16 | -0.61 |
| Crystallite size (nm) | -- | 114 | 101 | 19 | 104 | 41 |

Table. 2 Raman peak properties (position ($\varpi$), full width half maximum ($\beta$), and Normalized Intensity (N.I.)) of Eu doped M-type barium hexaferrite samples.

| Peak No. | BM | | | B09E01 | | | B08E02 | | | B07E02 | | | B06E02 | | | Modes | Assigned polyhedral |
|---|---|---|---|---|---|---|---|---|---|---|---|---|---|---|---|---|---|
| | $\varpi$ ($cm^{-1}$) | $\beta$ ($cm^{-1}$) | N. I. | $\varpi$ ($cm^{-1}$) | $\beta$ ($cm^{-1}$) | N. I. | $\varpi$ ($cm^{-1}$) | $\beta$ ($cm^{-1}$) | N. I. | $\varpi$ ($cm^{-1}$) | $\beta$ ($cm^{-1}$) | N.I. | $\varpi$ ($cm^{-1}$) | $\beta$ ($cm^{-1}$) | N.I. | | |
| 1 | 90.01(6) | 5.84(3) | 0.50 | 89.74(1) | 7.19(9) | 0.46 | 89.46(4) | 7.27(4) | 0.57 | 89.64(2) | 7.26(9) | 0.24 | 89.64(2) | 7.27(1) | 0.24 | $E_{1u}$ | Out-of-plane vibration of $Ba^{2+}$ ions |
| 2 | 172.91(2) | 6.91(1) | 0.25 | 173.01(1) | 9.09(6) | 0.28 | 172.60(8) | 9.72(4) | 0.33 | 172.97(7) | 8.26(7) | 0.13 | 172.97(9) | 8.26(1) | 0.13 | $E_{1g}$ | Motion of whole spinel block |
| 3 | 184.59(8) | 8.03(2) | 0.42 | 185.04(7) | 8.02(5) | 0.39 | 184.49(3) | 9.65(8) | 0.45 | 184.80(6) | 8.80(5) | 0.20 | 184.80(6) | 8.79(6) | 0.20 | $E_{1g}$ | |
| 4 | 213.75(2) | 9.50(3) | 0.10 | 214.02(8) | 8.90(7) | 0.10 | 213.42(4) | 8.38(2) | 0.12 | 214.38(1) | 8.80(9) | 0.06 | 214.38(3) | 8.78(9) | 0.06 | $E_{1g}$ | O-Fe-O vibrations |
| 5 | -- | -- | -- | 227.20(1)* | 7.38(1)* | 0.34* | 226.53(1)* | 8.27(5)* | 0.46* | 227.23(2)* | 7.56(6)* | 0.61* | 227.23(2)* | 7.56(1)* | 0.61* | $A_{1g}$* | *Movement of Fe-ions along c-axis in $Fe_2O_3$ |
| 6 | -- | -- | -- | 247.31(4)* | 5.73(6)* | 0.03* | 246.03(9)* | 5.73(9)* | 0.04* | 245.85(5)* | 7.47(3)* | 0.06* | 245.87(9)* | 7.37(7)* | 0.06* | $E_g$* | *Rotations and translations of $Fe_2O_3$ |
| 7 | 285.73(5) | 15.10(7) | 0.16 | 293.60(8)* | 9.60(5)* | 0.61* | 293.36(4)* | 8.74(1)* | 0.92* | 294.62(3)* | 8.18(1)* | 1.00* | 294.70(3)* | 8.11(4)* | 1.00* | $E_{1g}/E_g$* | *Rotations and translations of $Fe_2O_3$ |
| 8 | 317.35(5) | 11.47(3) | 0.11 | 318.66(4) | 11.73(9) | 0.16 | 317.05(9) | 15.19(5) | 0.16 | 317.75(6) | 5.3895) | 0.06 | 318.81(3) | 9.18(1) | 0.06 | $A_{1g}$ | |
| 9 | 337.97(2) | 11.62(1) | 1.00 | 338.93(3) | 15.13(1) | 1.00 | 338.58(1) | 17.56(3) | 0.92 | 339.03(2) | 8.83(1) | 0.45 | 338.89(1) | 8.17(9) | 0.46 | $A_{1g}$ | Octahedral (mixed) vibrations |
| 10 | 384.85(7) | 6.20(9) | 0.05 | 385.51(8) | 10.29(9) | 0.08 | 384.90(2) | 9.66(3) | 0.05 | 385.28(2) | 8.87(9) | 0.03 | 385.30(6) | 8.90(1) | 0.03 | $E_{2g}$ | |
| 11 | 410.17(0) | 10.27(2) | 0.14 | -- | -- | -- | -- | -- | -- | -- | -- | -- | -- | -- | -- | $A_{1g}$ | Fe atom vibrations at the 12k sites |
| 12 | 420.36(7) | 13.60(1) | 0.22 | 415.06(1)* | 18.69(5)* | 0.48* | 413.63(3)* | 19.18(8)* | 0.58* | 413.01(6)* | 15.68(4)* | 0.56* | 413.02(1)* | 15.71(1)* | 0.56* | $E_{2g}/E_g$* | |
| 13 | 452.96(5) | 8.36(3) | 0.02 | 453.20(1) | 8.25(3) | 0.02 | 453.49(7) | 8.70(9) | 0.03 | -- | -- | -- | -- | -- | -- | $E_{1g}$ | |
| 14 | 469.62(6) | 13.68(9) | 0.15 | 470.08(1) | 12.54(1) | 0.15 | 469.65(4) | 15.64(9) | 0.16 | 470.28(1) | 11.23(4) | 0.06 | 470.28(1) | 11.19(9) | 0.06 | $A_{1g}$ | Fe atom vibrations at 2a & 12k sites |
| 15 | -- | -- | -- | 501.02(8)* | 8.81(4)* | 0.02* | 506.47(2)* | 30.35(9)* | 0.04* | 499.40(9)* | 15.31(9)* | 0.05* | 499.39(4)* | 15.25(1)* | 0.05* | $A_{1g}$* | |
| 16 | 528.90(2) | 9.09(1) | 0.44 | 529.58(2) | 18.3(2) | 0.51 | 529.05(9) | 18.49(8) | 0.45 | 529.66(1) | 17.26(6) | 0.21 | 529.66(1) | 17.27(2) | 0.22 | $E_{2g}$ | |
| 17 | 564.17(4) | 8.17(9) | 0.01 | 566.65(3) | 0.65(2) | 0.01 | -- | -- | -- | -- | -- | -- | -- | -- | -- | $E_{2g}$ | |
| 18 | 615.86(8) | 22.24(4) | 0.32 | 615.11(1) | 22.05(6) | 0.44 | 614.81(8) | 23.36(2) | 0.46 | 613.95(3) | 17.23(9) | 0.35 | 613.95(3) | 17.25(9) | 0.35 | $A_{1g}$ | Vibrations of O atoms at $4f_2$ sites |
| 19 | 685.07(8) | 7.88(5) | 0.96 | 685.83(9) | 20.78(4) | 0.96 | 685.58(9) | 22.96(2) | 1.00 | 685.68(6) | 13.33(2) | 0.45 | 685.68(6) | 13.33(3) | 0.45 | $A_{1g}$ | Vibrations of O atoms at 2b sites |
| 20 | 718.18(8) | 22.73(2) | 0.20 | 718.60(8) | 20.69(7) | 0.22 | 718.16(4) | 22.13(2) | 0.21 | 718.15(3) | 21.06(2) | 0.10 | 718.15(3) | 21.06(4) | 0.10 | $A_{1g}$ | Vibrations of O atoms at $4f_1$ sites |

Note: Hematite phase modes are indicated with * symbol

Table. 3 Summary of the peak parameters (Area, position, and FWHM) from deconvolution of the XPS spectra and fitted with PseduoVoigt function.

| | | BM | | | B09E01 | | | B08E02 | | | B10E02 | | | B07E02 | | | B06E02 | | |
|---|---|---|---|---|---|---|---|---|---|---|---|---|---|---|---|---|---|---|---|
| Core Bands | Peak No. | A (arb. units) | B.E. (eV) | β (eV) | A (arb. units) | B.E. (eV) | β (eV) | A (arb. units) | B.E. (eV) | β (eV) | A (arb. units) | B.E. (eV) | β (eV) | A (arb. units) | B.E. (eV) | β (eV) | A (arb. units) | B.E. (eV) | β (eV) |
| O1s Scan | 1 | 205408.18 | 529.56(10) | 1.45(1) | 368269.86 | 529.84(6) | 1.54(13) | 250413.93 | 529.69(5) | 1.43(3) | 322379.85 | 529.78(91) | 1.60(1) | 266176.69 | 529.72(8) | 1.45(3) | 210471.09 | 529.71(6) | 1.31(8) |
| | 2 | 106541.28 | 530.97(20) | 2.26(43) | 95910.40 | 531.07(72) | 2.77(40) | 113080.92 | 530.67(18) | 2.49(29) | 32752.17 | 531.40(23) | 1.14(28) | 104578.94 | 530.84(8) | 2.34(51) | 111540.17 | 531.02(8) | 3.36(7) |
| | 3 | 34747.79 | 532.52(27) | 2.17(48) | 53696.40 | 531.89(40) | 1.81(32) | 29368.86 | 532.00(17) | 1.94(38) | 52010.34 | 531.94(8) | 1.67(96) | 31068.93 | 532.13(11) | 1.82(16) | 167762.84 | 531.69(8) | 1.78(8) |
| Ba 3d Scan | | | | | | | | | | | | | | | | | | | |
| Ba $3d_{5/2}$ | 1 | 100430.54 | 778.76(3) | 1.28(2) | 166040.97 | 779.15(3) | 1.52(2) | 98853.77 | 778.84(2) | 1.21(1) | 143879.11 | 778.93(3) | 1.41(2) | 94554.59 | 778.92(4) | 1.31(2) | 134868.51 | 779.18(11) | 2.17(7) |
| | 2 | 196758.70 | 779.84(1) | 1.74(1) | 112413.26 | 780.14(4) | 1.83(4) | 123768.64 | 779.86(2) | 1.73(1) | 156237.12 | 780.02(2) | 1.59(2) | 139057.88 | 779.85(2) | 1.96(6) | 171061.45 | 780.26(8) | 1.44(6) |
| Ba $3d_{3/2}$ | 3 | 90250.65 | 794.22(3) | 1.45(2) | 106695.20 | 794.36(4) | 1.32(2) | 60531.33 | 794.12(4) | 1.16(2) | 119208.07 | 794.31(3) | 1.54(2) | 46342.65 | 794.09(6) | 1.06(3) | 70213.82 | 794.35(16) | 1.53(8) |
| | 4 | 105899.32 | 795.32(2) | 1.51(2) | 85458.01 | 795.47(4) | 1.54(3) | 96224.16 | 795.13(2) | 1.79(2) | 74023.82 | 795.53(4) | 1.43(3) | 103239.23 | 795.03(2) | 1.89(2) | 127483.05 | 795.74(7) | 1.54(5) |
| Eu 3d Scan | | | | | | | | | | | | | | | | | | | |
| Eu $3d_{5/2}$ | 1 | -- | -- | -- | 109743.87 | 1134.58(5) | 4.60(6) | 108967.33 | 1134.35(5) | 4.51(5) | 91466.57 | 1134.15(5) | 4.86(5) | 130370.04 | 1135.04(5) | 4.61(5) | 96770.60 | 1134.89(5) | 4.69(5) |
| Eu $3d_{3/2}$ | 2 | -- | -- | -- | 52630.83 | 1164.26(13) | 5.24(18) | 45458.97 | 1164.10(13) | 4.49(11) | 33773.60 | 1163.79(14) | 4.30(17) | 74429.85 | 1164.56(11) | 5.04(15) | 21969.19 | 1164.31(16) | 3.86(19) |
| Fe 2p Scan | | | | | | | | | | | | | | | | | | | |
| Fe $2p_{3/2}$ | 1 | 301836.48 | 710.80(56) | 3.09(24) | 138611.48 | 710.50(24) | 2.17(74) | 129161.50 | 710.25(23) | 2.08(8) | 174793.65 | 710.32(17) | 2.28(24) | 164170.43 | 710.25(26) | 2.12(25) | 187084.06 | 710.33(80) | 2.28(23) |
| | 2 | 332687.39 | 712.81(83) | 4.72(70) | 252286.36 | 711.73(33) | 3.99(11) | 149106.92 | 711.82(86) | 3.38(64) | 195480.11 | 711.87(2) | 3.51(13) | 214405.71 | 711.88(69) | 2.98(86) | 117366.56 | 712.22(50) | 2.53(70) |
| | 3 | 178184.05 | 714.55(55) | 4.00(12) | 111731.12 | 714.77(44) | 4.84(32) | 174316.18 | 713.12(19) | 5.10(24) | 152307.80 | 714.18(42) | 4.97(13) | 152056.88 | 714.44(99) | 3.91(90) | 122744.53 | 713.53(21) | 4.16(41) |
| Satellite | 4 | 323421.49 | 718.49(68) | 5.24(30) | 60147.29 | 718.99(98) | 3.31(14) | 124971.15 | 718.56(15) | 5.30(6) | 147566.49 | 718.91(19) | 4.39(46) | 123492.86 | 718.70(40) | 4.28(12) | 110853.82 | 718.50(18) | 5.37(47) |
| Fe $2p_{1/2}$ | 5 | 348803.72 | 724.71(3) | 5.32(6) | 124782.71 | 723.72(67) | 2.89(69) | 81836.55 | 724.28(64) | 3.12(45) | 154319.44 | 724.11(41) | 3.51(41) | 150479.66 | 723.80(48) | 3.47(55) | 96491.04 | 724.02(97) | 3.10(62) |
| | 6 | 172490.02 | 724.92(41) | 6.31(20) | 136538.47 | 725.43(17) | 4.21(59) | 131572.22 | 725.27(12) | 5.42(41) | 116245.64 | 726.06(8) | 5.56(34) | 148203.05 | 725.77(52) | 4.71(87) | 84279.87 | 726.11(50) | 4.62(43) |
| Satellite | 7 | 207096.41 | 730.55(33) | 8.60(36) | 23331.03 | 733.12(14) | 3.43(84) | 37933.79 | 732.39(77) | 5.63(39) | 42419.49 | 733.04(7) | 4.31(82) | 55047.51 | 732.74(6) | 5.53(3) | 31429.60 | 733.08(46) | 4.44(83) |

Table. 4 Ferrimagnetic parameters of Eu substituted $BaFe_{12}O_{19}$ samples

| Eu/Ba ratio | $\alpha$-$Fe_2O_3$ (%) | Samples | 3 K | | | | | 300 K | | | | |
|---|---|---|---|---|---|---|---|---|---|---|---|---|
| | | | $M_{Sat}$ (emu/g) | $M_R/M_{Sat}$ | $H_C$ (Oe) | $K_{eff}$ (M.Oe emu $g^{-1}$) | Loop area (Oe emu $g^{-1}$) | $M_{Sat}$ (emu/g) | $M_R/M_{Sat}$ | $H_C$ (Oe) | $K_{eff}$ (M.Oe emu $g^{-1}$) | Loop area (Oe emu $g^{-1}$) |
| 0 | 0.00 | BM | 84.44 | 0.33 | 914 | 0.451 | 188091.93 | 67.09 | 0.34 | 1996 | 2.489 | 294776.42 |
| 0.2 | 3.35 | B10E02 | 82.92 | 0.50 | 3657 | 1.086 | 696588.48 | 63.50 | 0.49 | 4863 | 2.400 | 657284.30 |
| 0.11 | 8.86 | B09E01 | 77.70 | 0.46 | 2815 | 1.055 | 486142.11 | 60.43 | 0.46 | 3966 | 2.320 | 481940.80 |
| 0.25 | 13.90 | B08E02 | 66.65 | 0.50 | 3632 | 1.378 | 550918.29 | 52.81 | 0.48 | 4802 | 2.180 | 535611.36 |
| 0.29 | 20.07 | B07E02 | 58.28 | 0.48 | 3418 | 1.279 | 451376.77 | 44.55 | 0.48 | 4584 | 1.814 | 426452.50 |
| 0.33 | 26.39 | B06E02 | 57.68 | 0.25 | 552 | 1.445 | 89470.27 | 43.42 | 0.28 | 1305 | 1.801 | 109704.39 |

Table. 5 Electrical parameters obtained from analysis of the σ(T) at 200 K and J(E) curves at 300 K.

| Eu/Ba ratio | Samples | σ(T) analysis at 200 K | | | | | J(E) curves analysis at 300 K | | | | |
|---|---|---|---|---|---|---|---|---|---|---|---|
| | | MVRH parameters | | | | | Conductivity (S/m) @ 20 kV/m | Poole-Frenkel emission | | Schottky emission | |
| | | $T_o$ (K) | ξ (Å) | $N(E_F)$ ($eV^{-1}m^{-3}$) | $R_H$ (nm) | $W_H$ (eV) | | $\varphi_T$ $(eV)$ | $\varepsilon_r$ | $\varphi_B$ $(eV)$ | $\varepsilon_r$ ($10^{-3}$) |
| 0.00 | BM | $2.68 \times 10^{9}$ | 3.9857 | $1.64 \times 10^{24}$ | 4.87 | 0.26 | $4.74 \times 10^{-6}$ | 0.33 | 0.41 | 0.72 | 4.87 |
| 0.11 | B09E01 | $3.11 \times 10^{9}$ | 3.9789 | $1.42 \times 10^{24}$ | 5.04 | 0.27 | $2.81 \times 10^{-5}$ | 0.28 | 1.57 | 0.66 | 8.60 |
| 0.20 | B10E02 | $4.12 \times 10^{4}$ | 3.9798 | $1.07 \times 10^{29}$ | 0.30 | 0.02 | $1.10 \times 10^{-5}$ | 0.31 | 0.44 | 0.74 | 1.50 |
| 0.25 | B08E02 | $3.94 \times 10^{9}$ | 3.9785 | $1.12 \times 10^{24}$ | 5.35 | 0.29 | $3.16 \times 10^{-5}$ | 0.28 | 0.46 | 0.65 | 9.67 |
| 0.29 | B07E02 | $3.08 \times 10^{9}$ | 3.9780 | $1.44 \times 10^{24}$ | 5.03 | 0.27 | $4.33 \times 10^{-5}$ | 0.28 | 0.29 | 0.65 | 7.01 |
| 0.33 | B06E02 | $7.56 \times 10^{9}$ | 3.9769 | $5.86 \times 10^{23}$ | 6.29 | 0.34 | $4.55 \times 10^{-6}$ | 0.34 | 0.33 | 0.73 | 4.87 |

Table. 6 Magneto-resistance and fitting parameters of Eu doped $BaFe_{12}O_{19}$

| Eu/Ba ratio | Samples | MR (%) @ 8 kOe | | α | | β | | γ | |
|---|---|---|---|---|---|---|---|---|---|
| | | E ∥ H | E ⊥ H | E ∥ H | E ⊥ H | E ∥ H | E ⊥ H | E ∥ H | E ⊥ H |
| 0.00 | BM | -37.59 | 3.13 | $3.27 \times 10^{-5}$ | $-1.23 \times 10^{-7}$ | $-4.22 \times 10^{-5}$ | $1.80 \times 10^{4}$ | 1.97 | -0.81 |
| 0.11 | B09E01 | -2.13 | -1.42 | $1.75 \times 10^{-7}$ | $1.09 \times 10^{-7}$ | $-1.55 \times 10^{-6}$ | $-3.89 \times 10^{-7}$ | 1.78 | 1.88 |
| 0.20 | B10E02 | -51.36 | 33.12 | $4.61 \times 10^{-6}$ | $1.03 \times 10^{-7}$ | $-3.44 \times 10^{-5}$ | $2.09 \times 10^{-1}$ | 1.79 | 0.54 |
| 0.25 | B08E02 | -3.96 | -1.91 | $2.39 \times 10^{-8}$ | $-1.41 \times 10^{-8}$ | $-1.57 \times 10^{4}$ | $-2.47 \times 10^{-2}$ | -0.86 | 0.42 |
| 0.29 | B07E02 | -7.06 | -1.84 | $3.52 \times 10^{-8}$ | $-9.71 \times 10^{-9}$ | $-2.89 \times 10^{-2}$ | $-2.18 \times 10^{-2}$ | 0.64 | 0.44 |
| 0.33 | B06E02 | -26.109 | -3.29 | $-9.01 \times 10^{-7}$ | $-4.12 \times 10^{-6}$ | $4.79 \times 10^{-13}$ | $2.44 \times 10^{-6}$ | 3.54 | 2.06 |

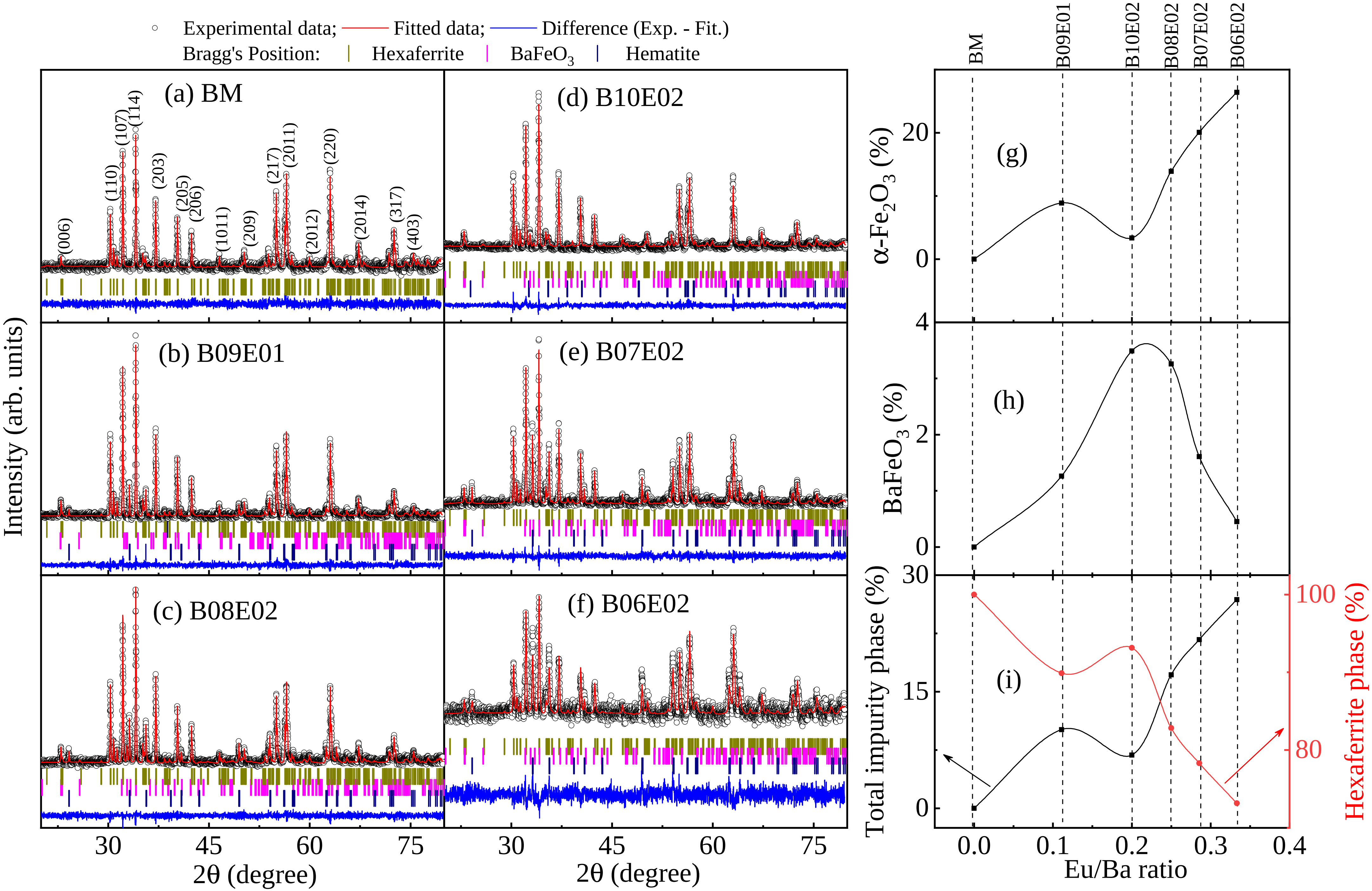

Fig.1 Rietveld refinement of Eu doped Barium hexaferrite XRD patterns (a-f). Variation of total impurity phase fraction and hexaferrite phase fraction (g), $\alpha$-$Fe_2O_3$ phase fraction (h), and $BaFeO_3$ phase fraction (i) with Eu/Ba ratio.

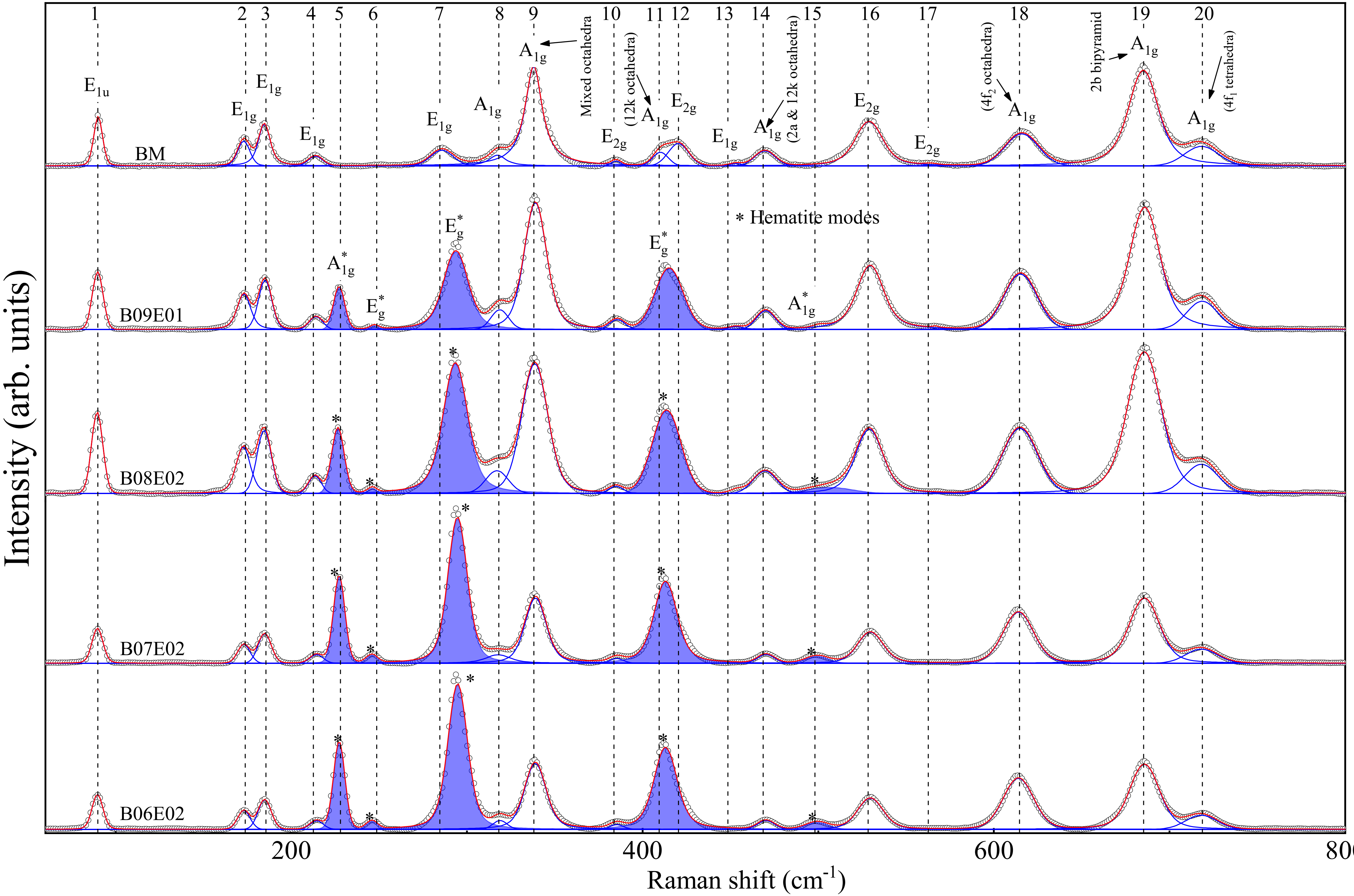

Fig. 2 Raman spectra of Eu substituted Barium hexaferrite samples. Note: Hematite modes indicate with * symbol.

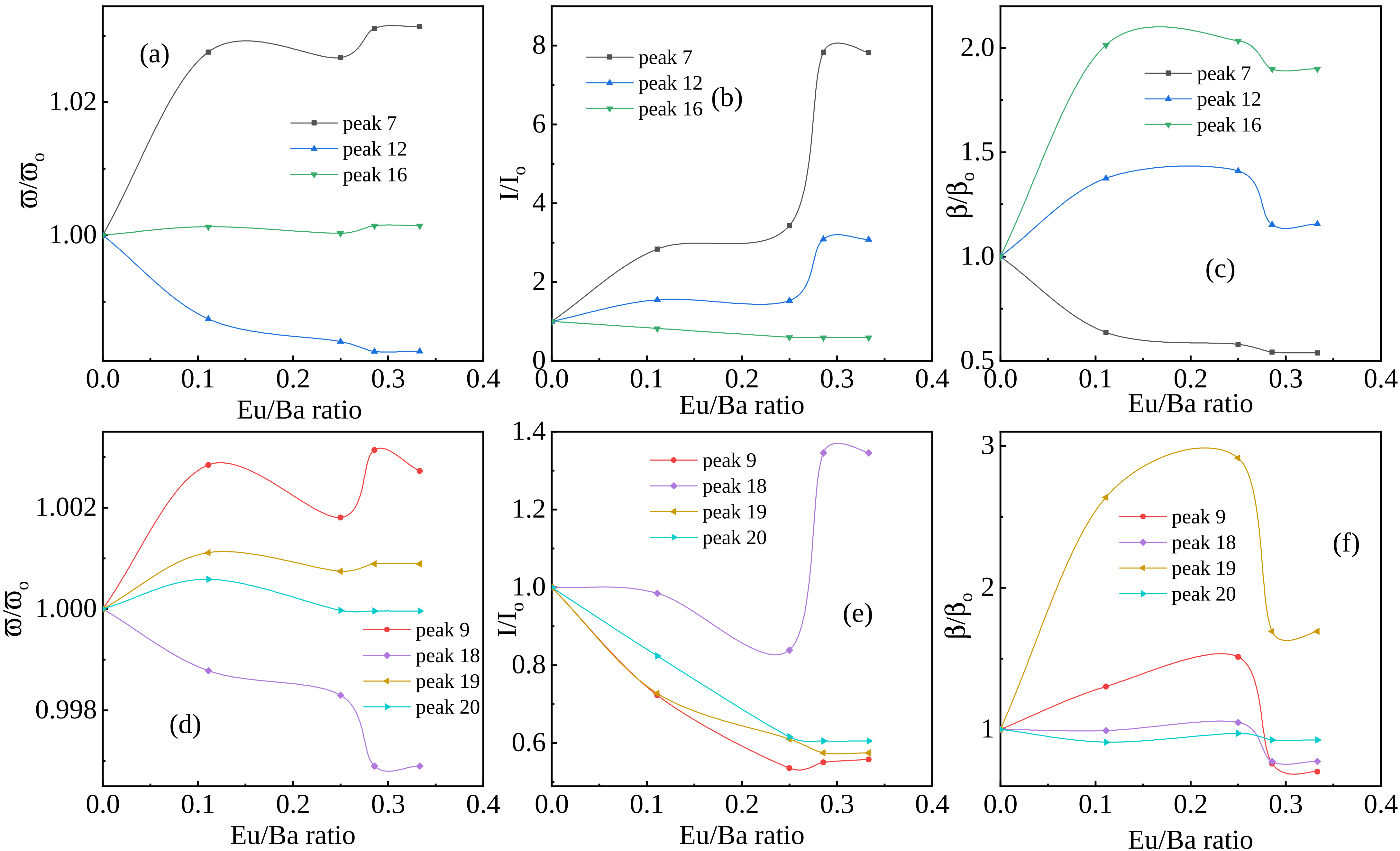

Fig. 3 Variation of normalized raman peak positions ($\varpi/\varpi_o$), Intensities ($I/I_o$), and FWHM ($\beta/\beta_o$) of in-plane vibrations (a-c) and out-of-plane vibrations (d-f) as function of Eu/Ba ratio in Eu substituted $BaFe_{12}O_{19}$.

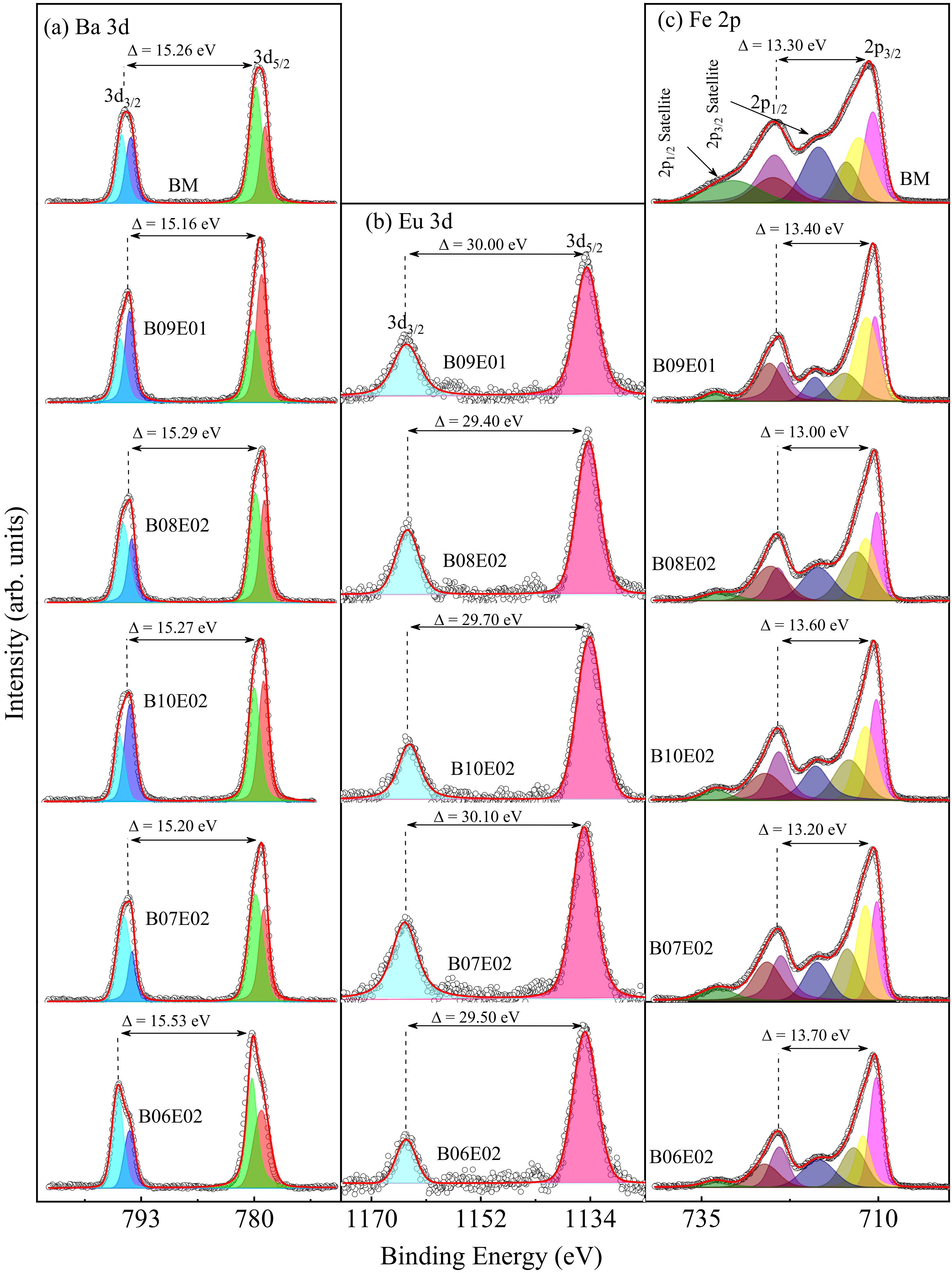

Fig. 4 XPS narrow scans of Ba 3d (a), Eu 3d (b), and Fe 2p (b) of Eu substituted Barium hexaferrite samples.

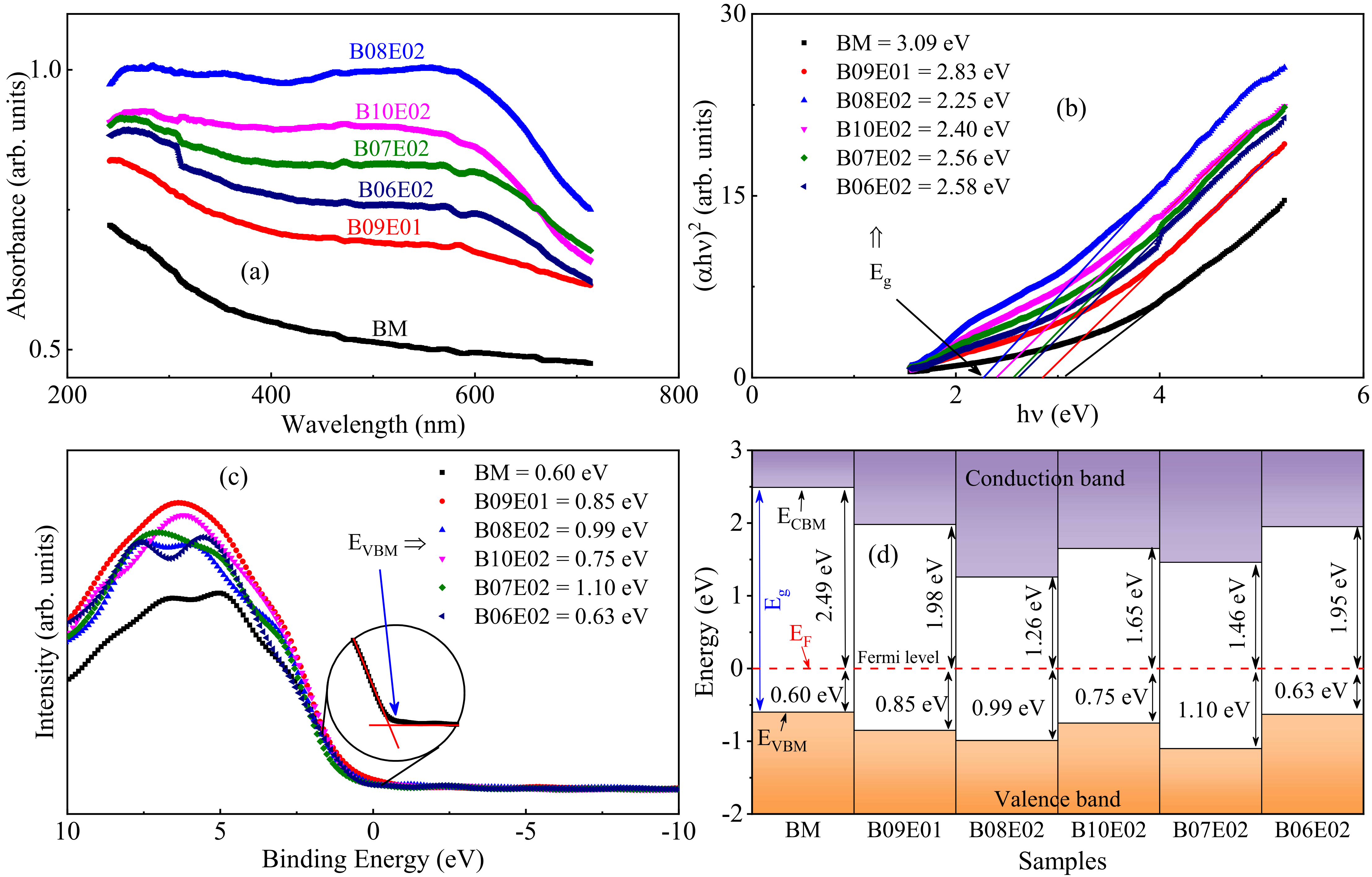

Fig. 5 Optical absorbance in the UV-Visible spectra (a), analysis using Tauc's plots (b), valence band spectra from XPS (c), and band structure diagram (d) for the Eu substituted Ba- hexaferrite samples.

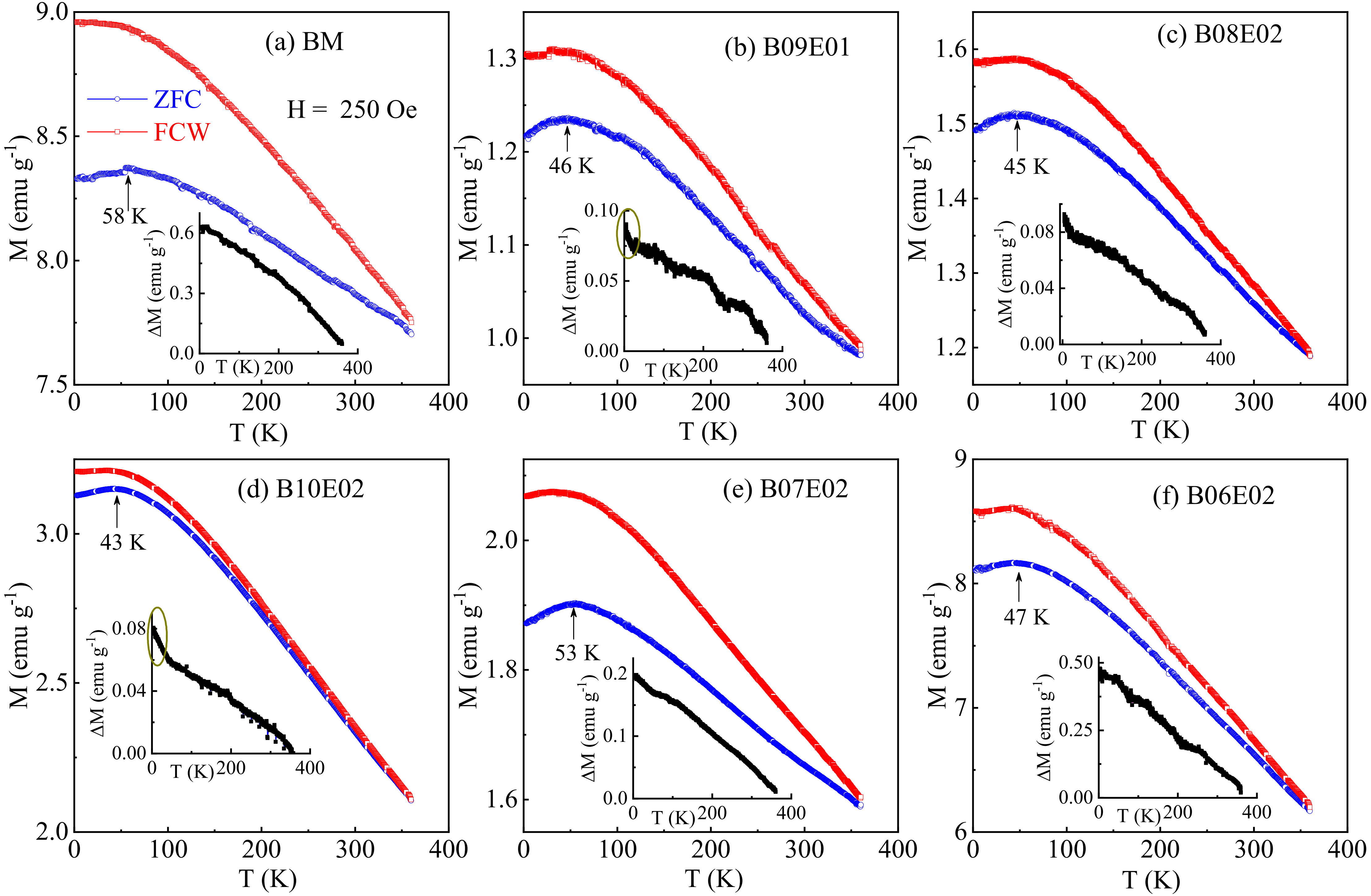

Fig. 6 The temperature dependence of magnetization curves under ZFC and FCW modes at 250 Oe for Eu substituted M-type Ba- hexaferrite samples. The insets show the temperature variation of the difference between the FC and ZFC magnetization.

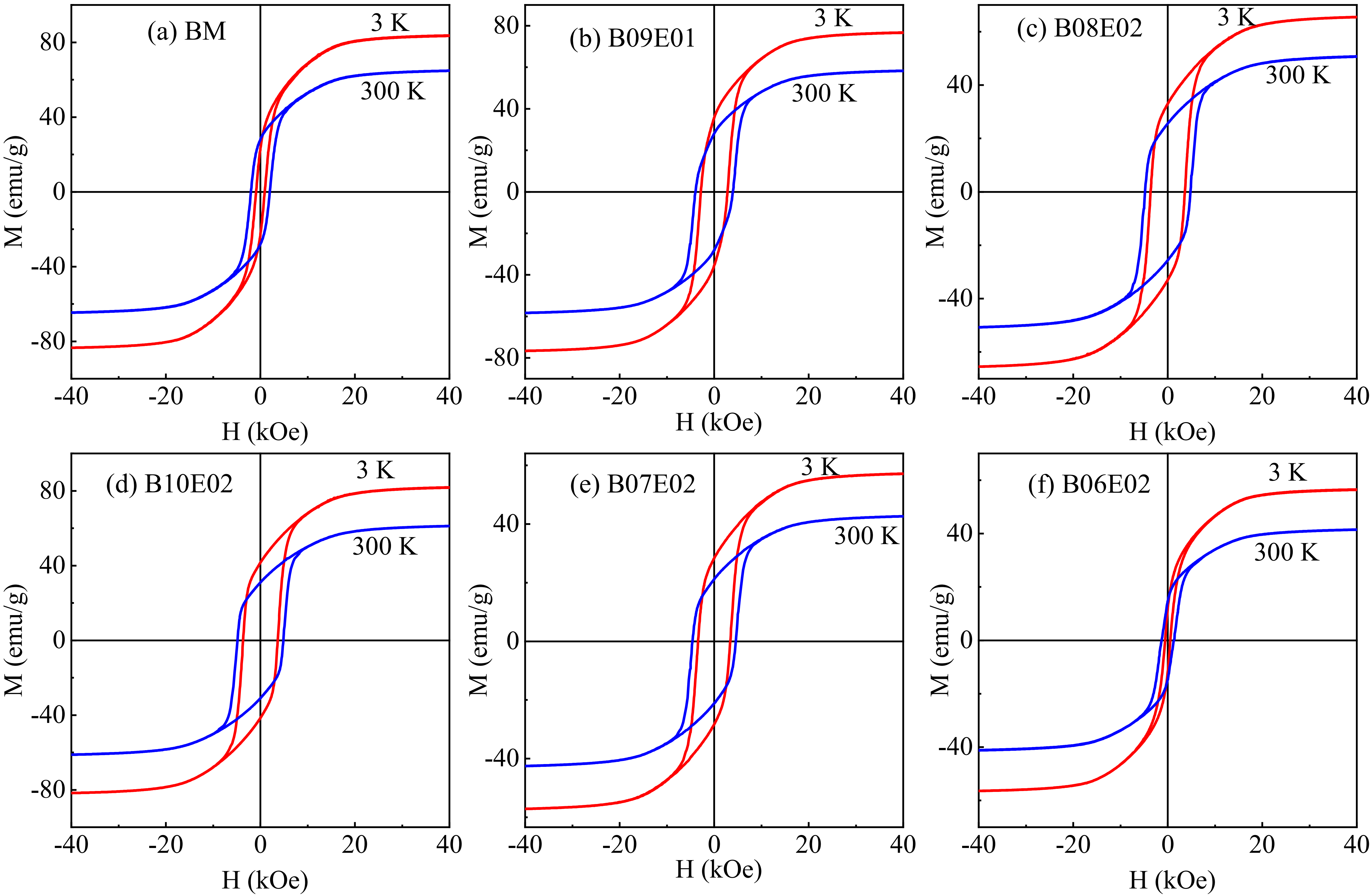

Fig. 7 M-H curves of Eu substituted M-type Barium hexaferrite at 3 K and 300 K.

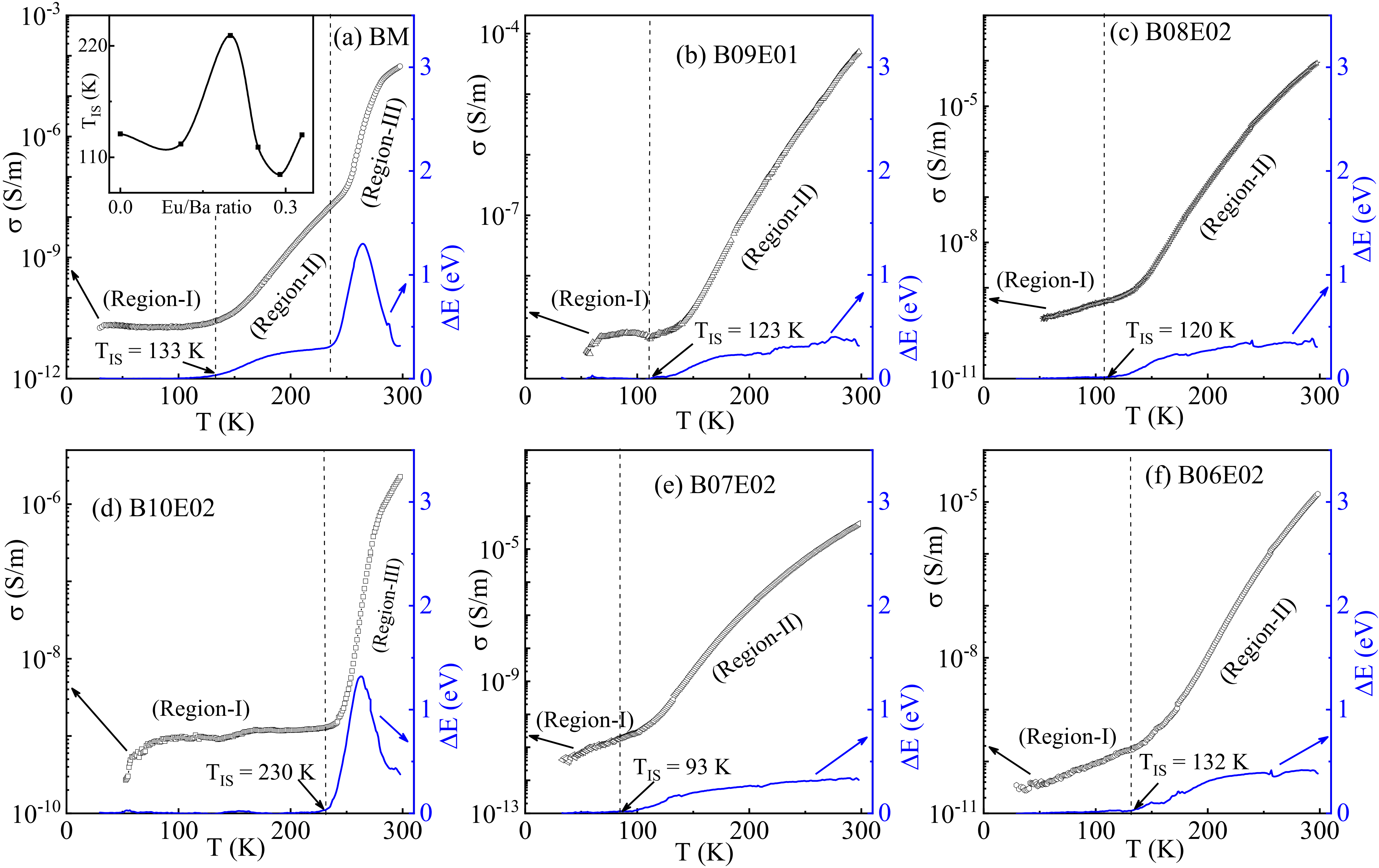

Fig. 8 Temperature dependence of the electrical conductivity (σ) (main panel-left Y axis), the local activation energy (ΔE) (main panel-right Y axis) and variation of the insulator-semiconductor transition temperature (inset of a) with Eu/Ba ratio for Eu substituted $BaFe_{12}O_{19}$ samples measured at electric field of 20 kV/m.

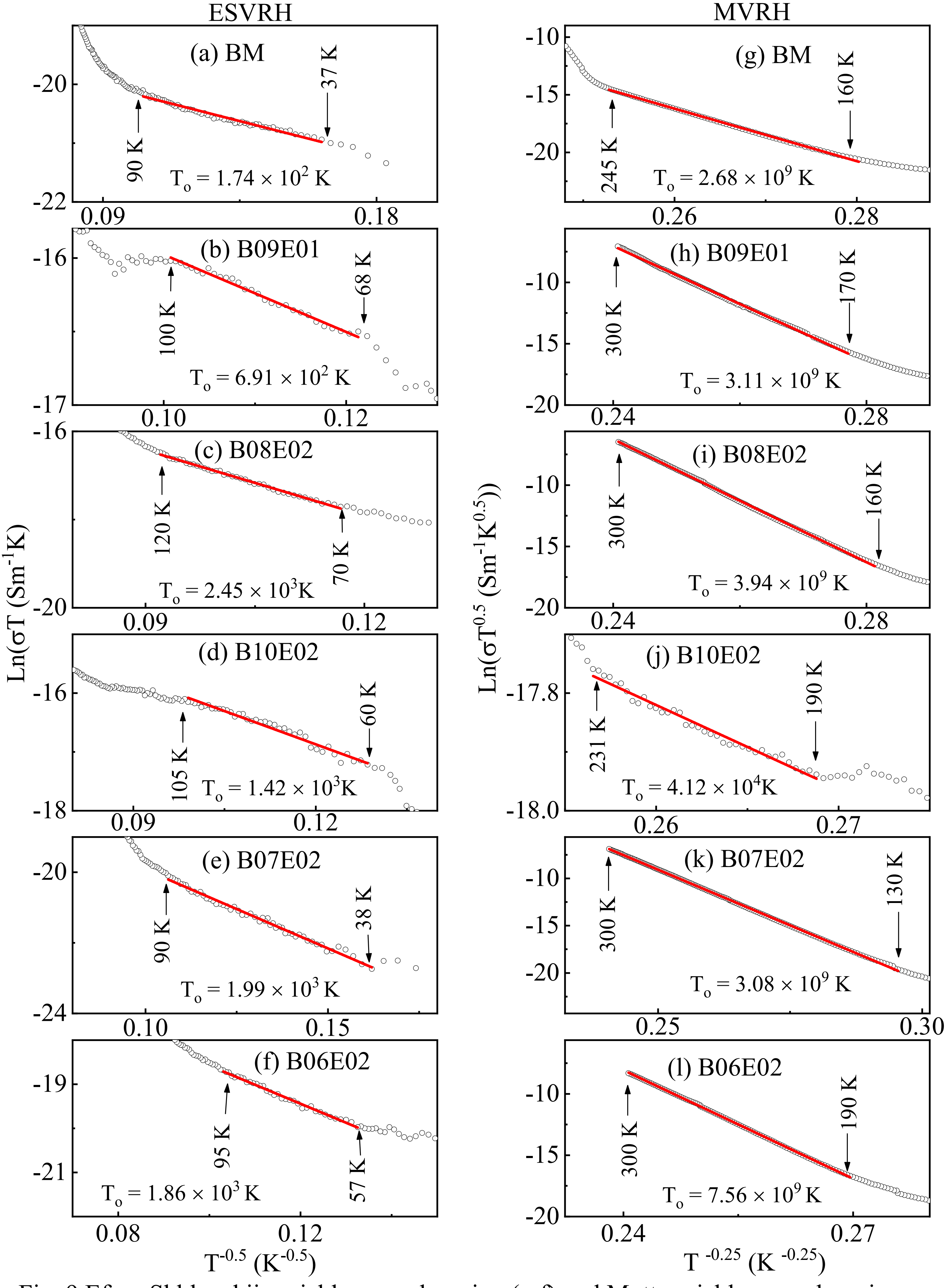

Fig. 9 Efros-Shklovskii variable range hopping (a-f) and Mott variable range hopping (g-l) mechanism of Eu substituted $BaFe_{12}O_{19}$.

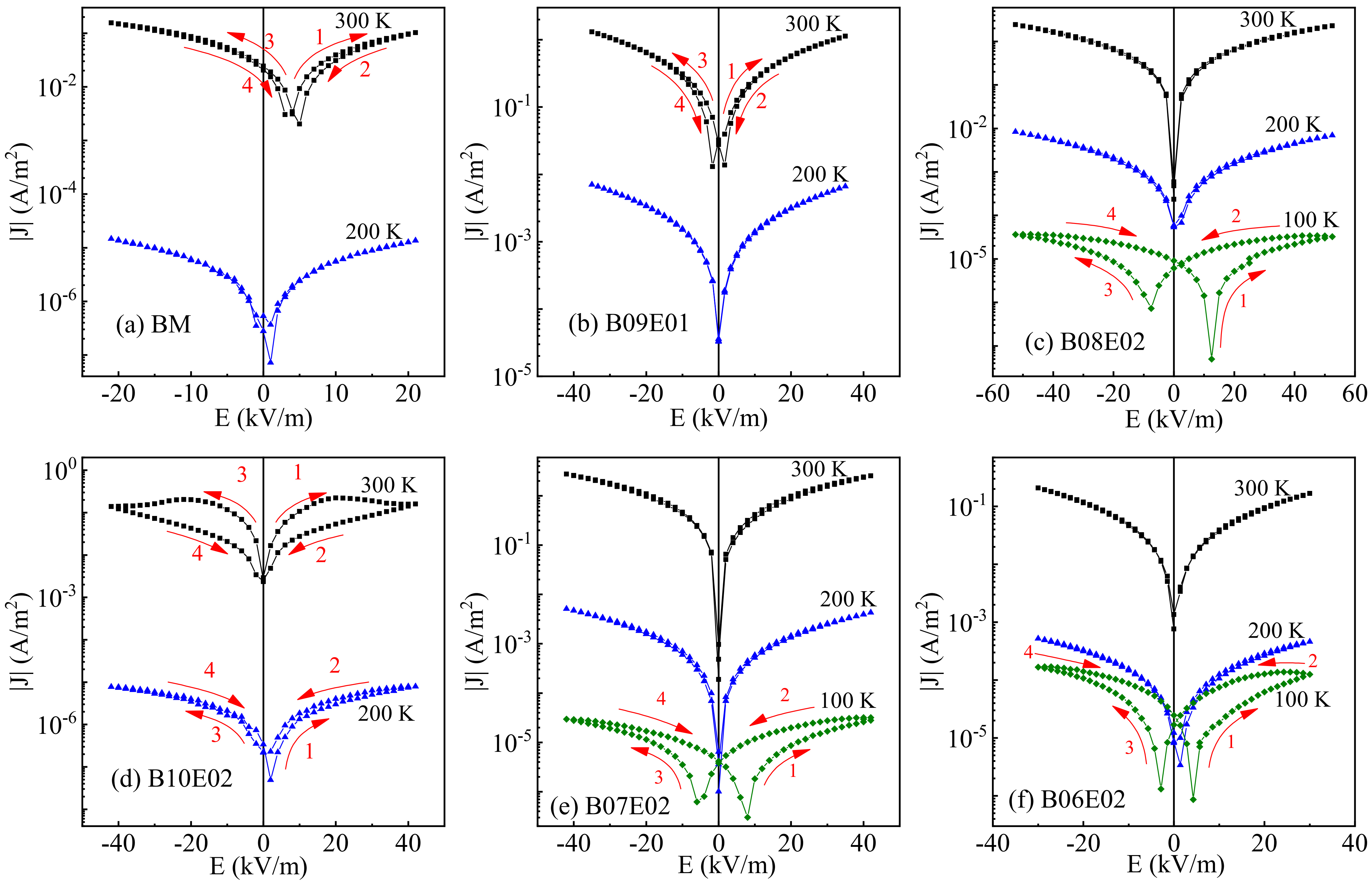

Fig. 10 J-E characteristics at selected temperatures for Eu substituted Ba-hexaferrite samples. The arrows indicate the electric field sweep directions.

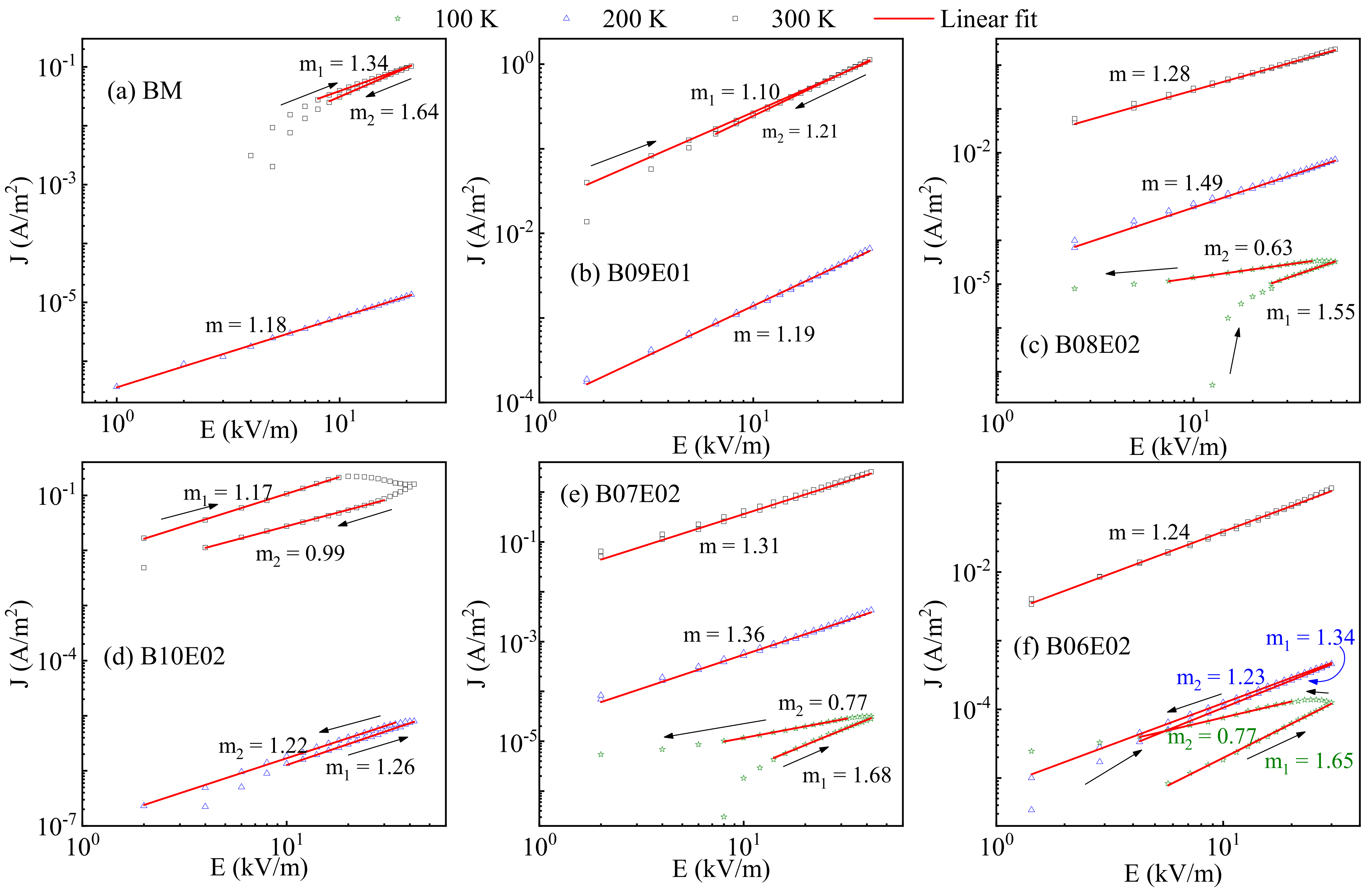

Fig. 11 Power law fit of J-E curves at 100 K, 200 K and 300 K. Note: $m_1$ and $m_2$ represent the slopes during forward and reverse paths of the resistive switching state, respectively. Otherwise, m represents the slope for the non-swiching state.

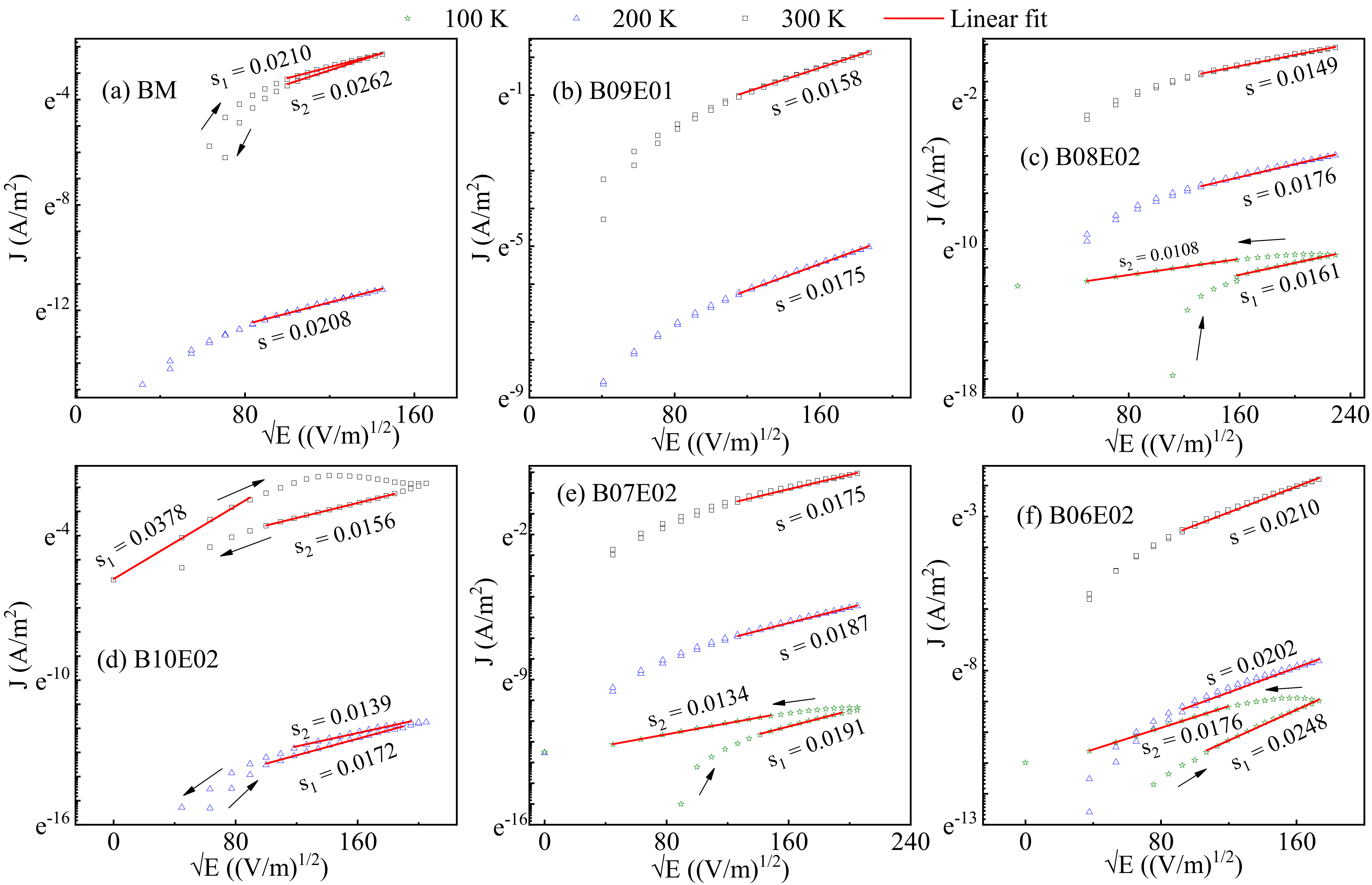

Fig. 12 Schottky emission mechanism is verified at 100 K, 200 K, and 300 K in the J-E curves for Eu substituted Ba-hexaferrite samples. Note: $s_1$ and $s_2$ are the slopes in the linear fits for forward and reverse paths, respectively.

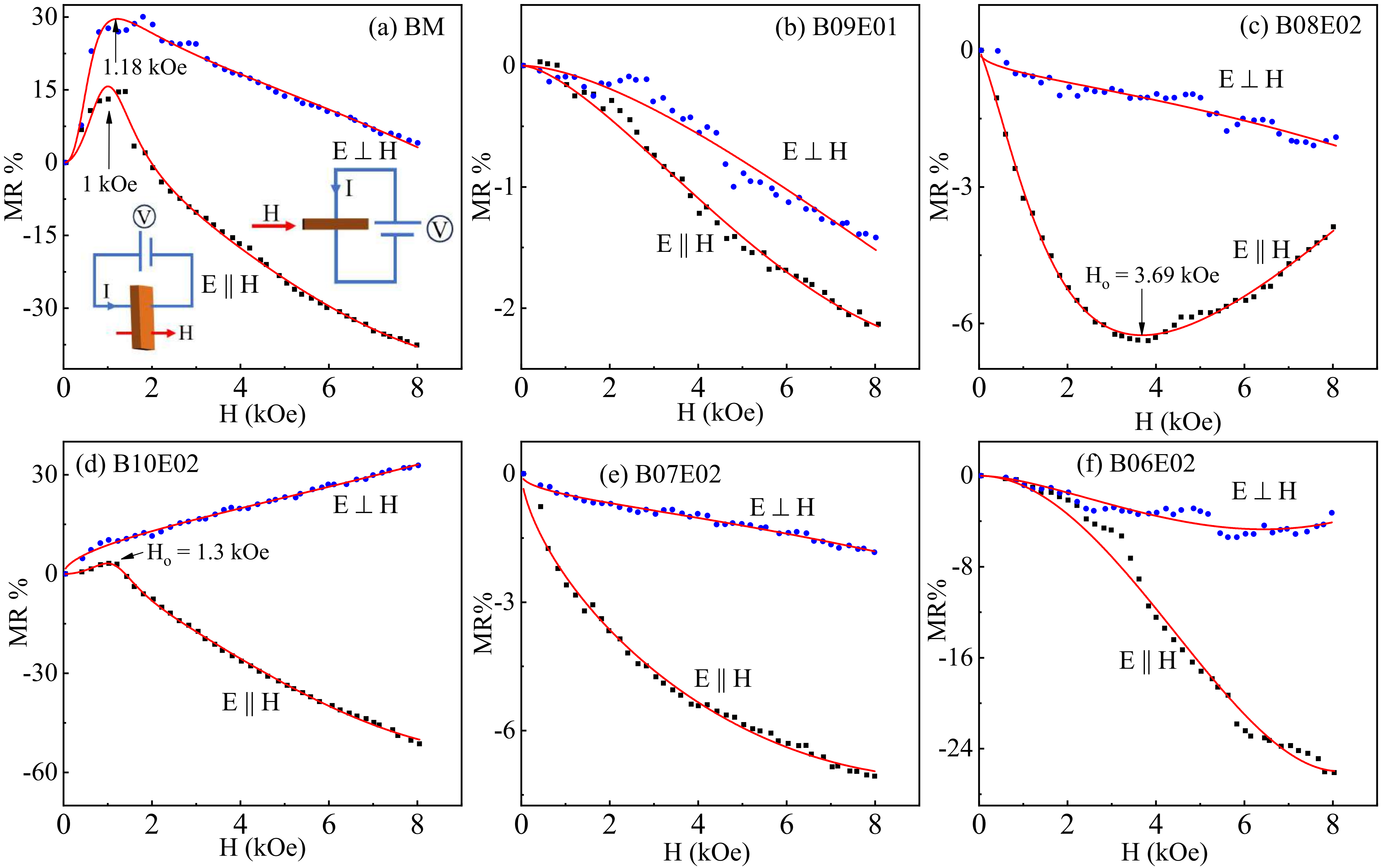

Fig. 13 Magnetic field dependence of MR values measured in longitudinal mode (E || H) and transverse mode (E ⊥ H) modes for the Eu substituted Ba-hexaferrite samples. The solid lines are the fitted data of the MR(H) curves.

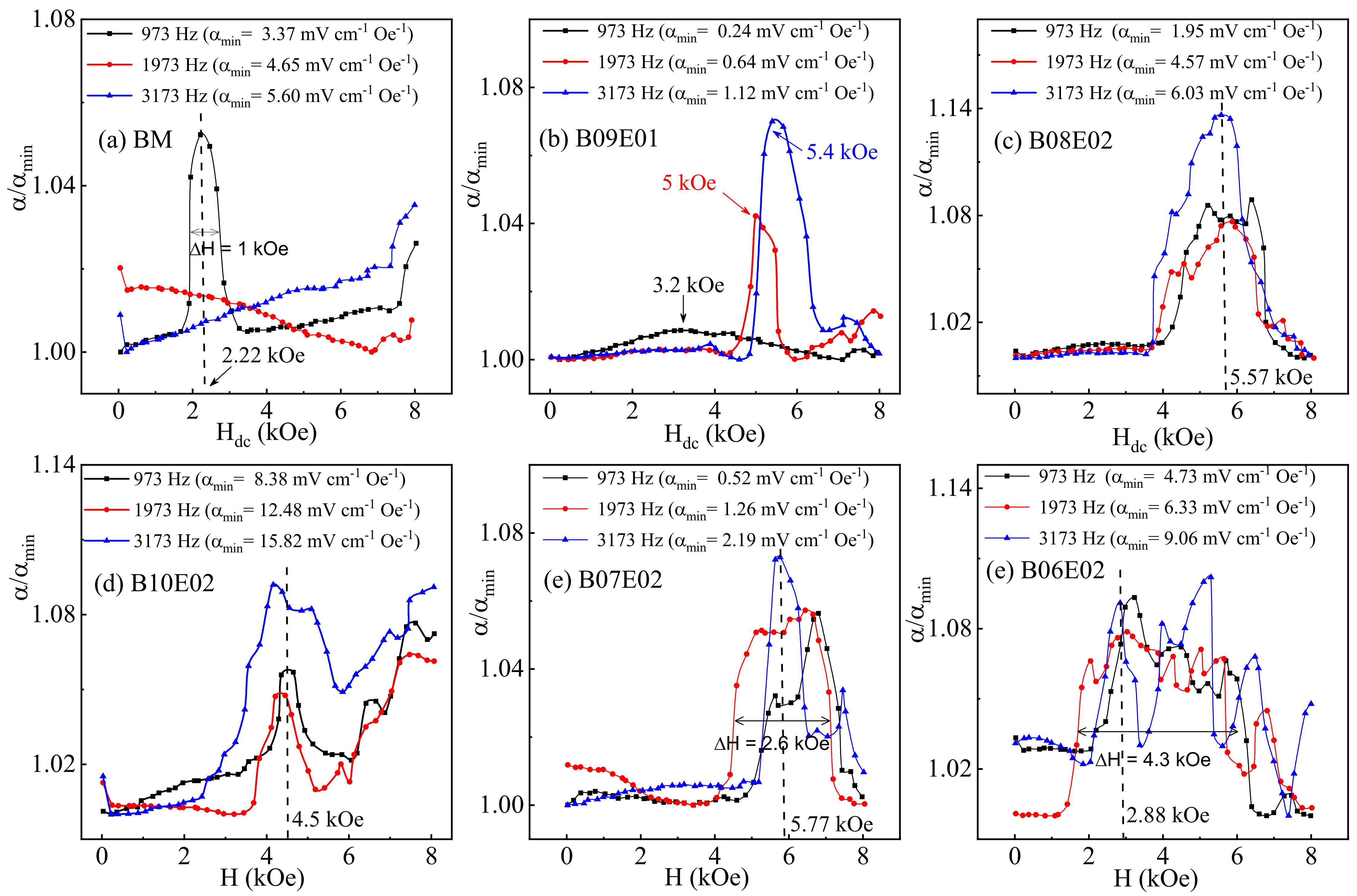

Fig. 14 The dc magnetic field dependence of the magneto-electric coupling constant in Eu substituted Ba-hexaferrite samples, measured by varying the frequency of the ac field amplitude at 4 Oe.

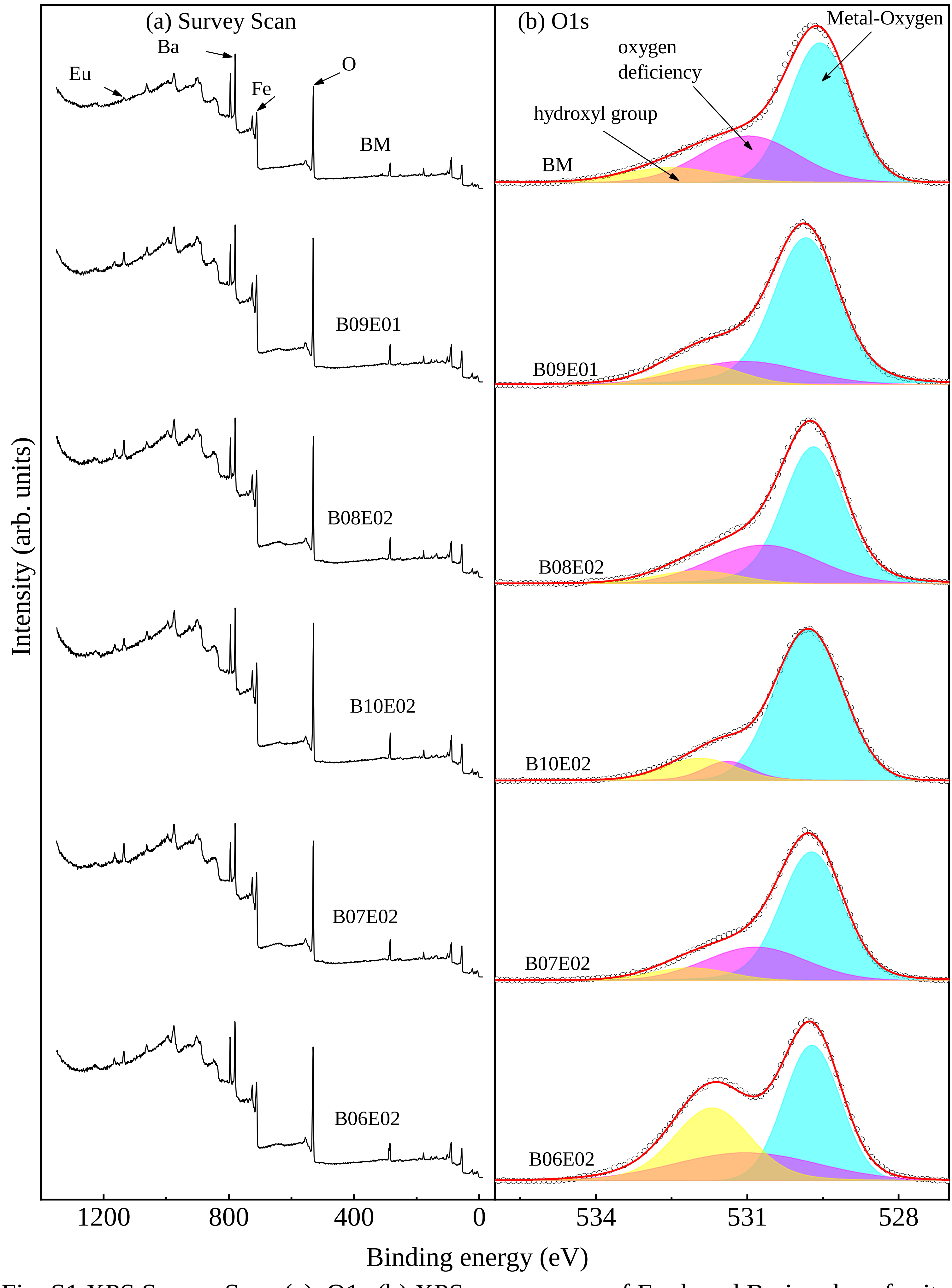

Fig. S1 XPS Survey Scan (a), O1s (b) XPS narrow scan of Eu doped Barium hexaferrite samples.

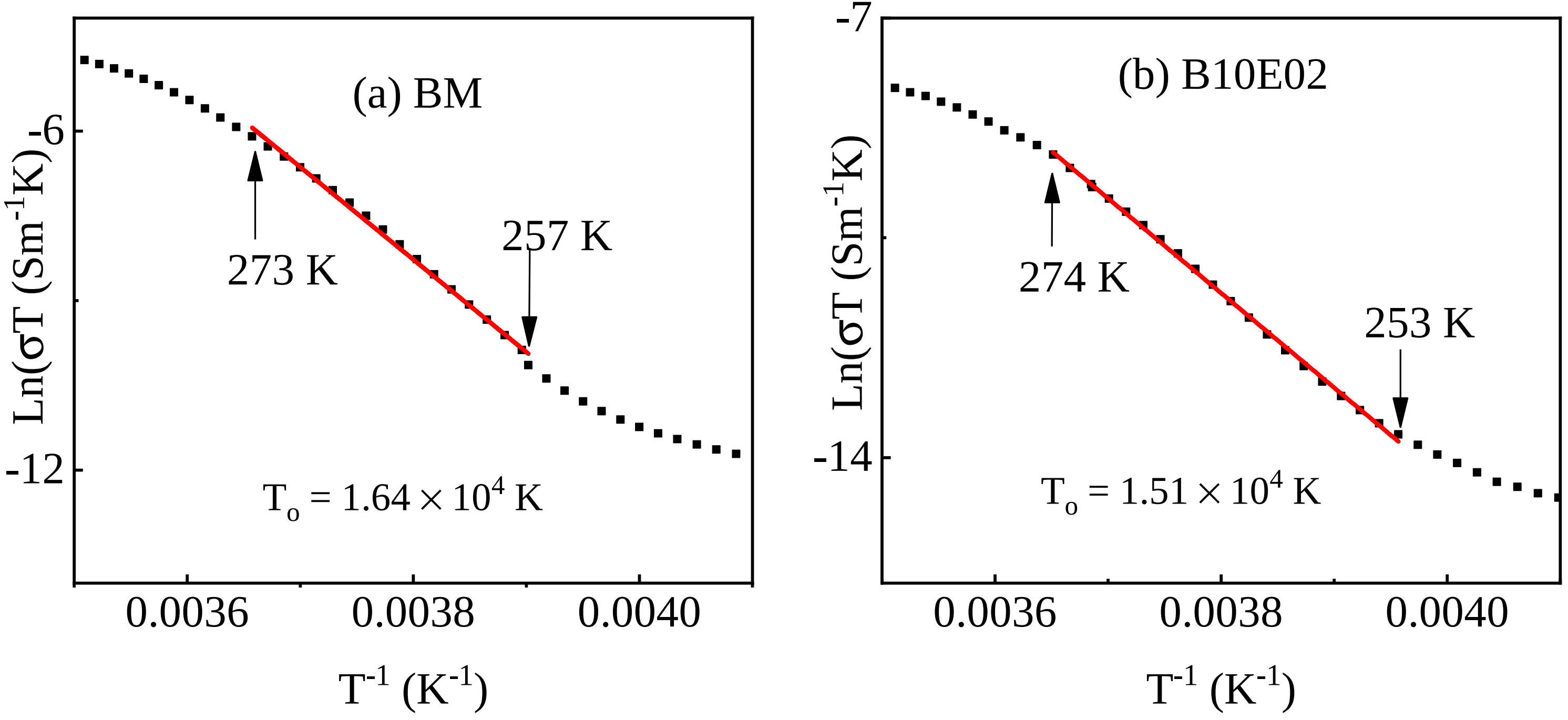

Fig. S2 Small polaron hopping mechanism of BM (a) and B10E02 (b) samples.

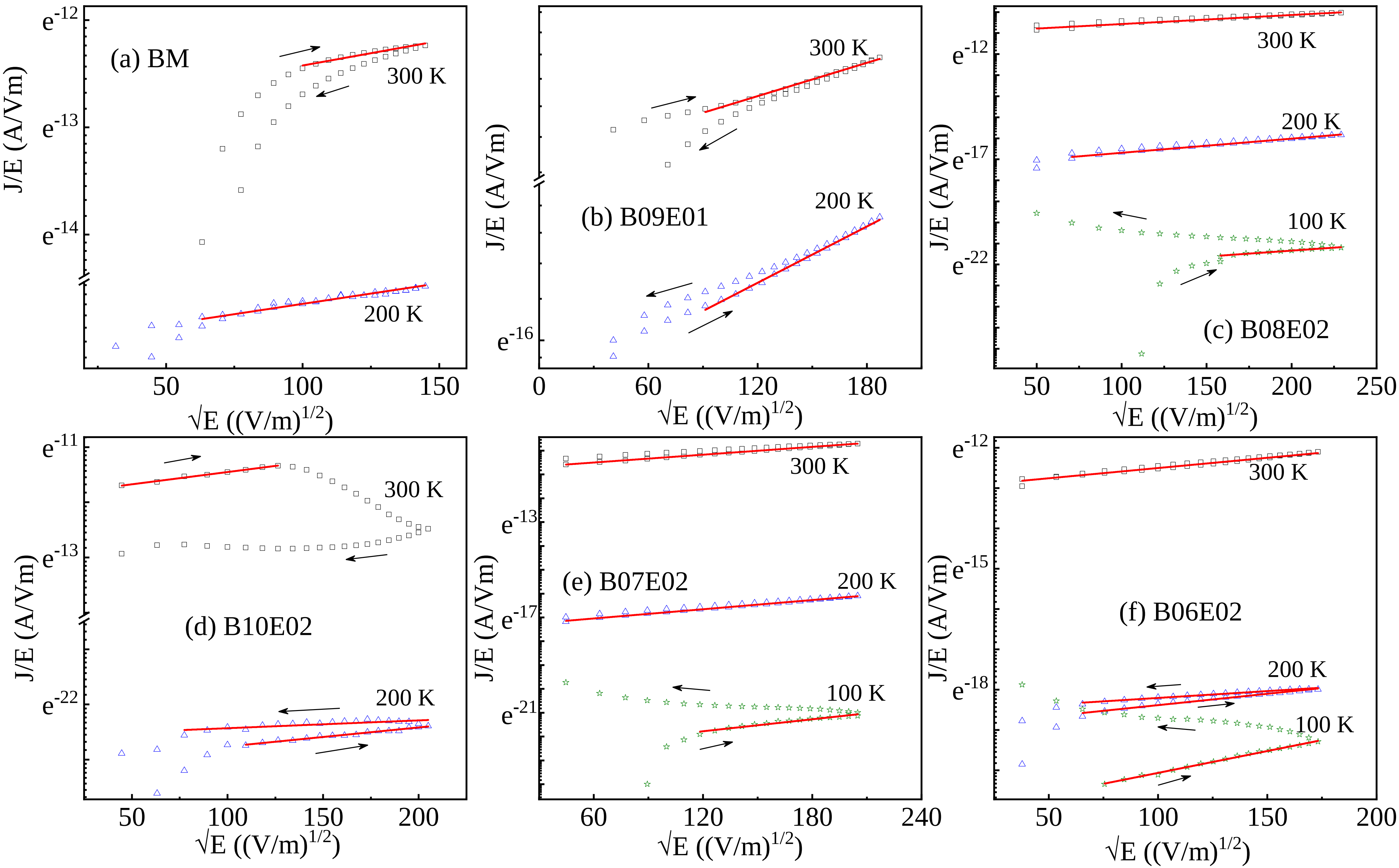

Fig. S3 Poole-Frenkel emission mechanism of J-E loops of Eu doped $BaFe_{12}O_{19}$ at various temperatures.